\documentclass[preprint2]{aastex62}
\usepackage{amsmath}
\usepackage{mathtools}
\usepackage{natbib}
\usepackage{threeparttable}
\usepackage{booktabs,tabulary}
\usepackage{bibentry}

\def\eg{{e.g.,}}
\def\ie{{\rm i.e.,}}

\newcommand{\be}{\begin{equation}}
\newcommand{\ee}{\end{equation}}

\newcommand{\flash}{{\sc FLASH}}
\newcommand{\maihem}{{\sc MAIHEM}}
\newcommand{\cloudy}{{\sc Cloudy}}

\newcommand{\HI}{\ion{H}{1}}
\newcommand{\HII}{\ion{H}{2}}
\newcommand{\HEI}{\ion{He}{1}}
\newcommand{\HEII}{\ion{He}{2}}
\newcommand{\HEIII}{\ion{He}{3}}
\newcommand{\CI}{\ion{C}{1}}

\newcommand{\CIII}{\ion{C}{3}}
\newcommand{\CIV}{\ion{C}{4}}

\newcommand{\CVII}{\ion{C}{7}}
\newcommand{\NI}{\ion{N}{1}}

\newcommand{\NVIII}{\ion{N}{8}}
\newcommand{\OI}{\ion{O}{1}}
\newcommand{\OII}{\ion{O}{2}}
\newcommand{\OIII}{\ion{O}{3}}

\newcommand{\OVI}{\ion{O}{6}}

\newcommand{\OIX}{\ion{O}{9}}
\newcommand{\NEI}{\ion{Ne}{1}}

\newcommand{\NEXI}{\ion{Ne}{11}}
\newcommand{\NAI}{\ion{Na}{1}}

\newcommand{\NAVI}{\ion{Na}{6}}
\newcommand{\MGI}{\ion{Mg}{1}}

\newcommand{\MGVI}{\ion{Mg}{6}}
\newcommand{\SiI}{\ion{Si}{1}}

\newcommand{\SiIII}{\ion{Si}{3}}
\newcommand{\SiIV}{\ion{Si}{4}}

\newcommand{\SiVI}{\ion{Si}{6}}
\newcommand{\SI}{\ion{S}{1}}

\newcommand{\SVI}{\ion{S}{6}}
\newcommand{\ARI}{\ion{Ar}{1}}

\newcommand{\ARIV}{\ion{Ar}{4}}

\newcommand{\ARVI}{\ion{Ar}{6}}
\newcommand{\CAI}{\ion{Ca}{1}}

\newcommand{\CAVI}{\ion{Ca}{6}}
\newcommand{\FEI}{\ion{Fe}{1}}

\newcommand{\FEVI}{\ion{Fe}{6}}
\newcommand{\ELEC}{${e}^{-}$}

\begin{document}

\revised{\today**}

\shorttitle{Line Emission from Cooling Superwinds}
\shortauthors{Gray et al.}

\title{Catastrophic Cooling in Superwinds: Line Emission and Non-equilibrium Ionization}
\author{William J. Gray}
\affil{Department of Astronomy, University of Michigan, 1085 South University Ave., Ann Arbor, Michigan 48109, USA}
\affil{CLASP, College of Engineering, University of Michigan, 2455 Hayward St., Ann Arbor, Michigan 48109, USA}
\author{M. S. Oey}
\affil{Department of Astronomy, University of Michigan, 1085 South University Ave., Ann Arbor, Michigan 48109, USA}
\author{Sergiy Silich}
\affil{Instituto Nacional de Astrof\'isica, Optica, y Electr\'onica, Puebla, AP 51, 72000 Puebla, Mexico}
\author{Evan Scannapieco}
\affil{School of Earth and Space Exploration, Arizona State University}

\begin{abstract}
Outflows are a pervasive feature of mechanical feedback from super star clusters (SSC) in starburst galaxies, playing a fundamental role in galaxy evolution. 
Observations are now starting to confirm that outflows can undergo catastrophic cooling, suppressing adiabatic superwinds.
Here we present a suite of one-dimensional, hydrodynamic simulations that study the ionization structure of these outflows and the resulting line emission generated by the cooling gas. 
We use the non-equilibrium atomic chemistry package within \maihem, our modified version of \flash, which evolves the ionization state of the gas and computes the total cooling rate on an ion-by-ion basis.
We find that catastrophically cooling models produce strong nebular line emission compared to adiabatic outflows.  We also show that such models exhibit non-equilibrium conditions, thereby generating more highly ionized states 
than equivalent equilibrium models.  
When including photoionization from the parent SSC, catastrophically cooling models show strong \CIV\ $\lambda 1549$ and \OVI\ $\lambda 1037$ emission.  For density bounded photoionization, \HEII\ $\lambda 1640$, $\lambda4686$, \CIII] $\lambda1908$, \SiIV\ $\lambda 1206$, and \SiIII\ $\lambda1400$ are also strongly enhanced.  These lines are seen in extreme starbursts where catastrophic cooling is likely to occur, suggesting that they may serve as diagnostics of such conditions.  The higher ionization generated by these flows may help to explain line emission that cannot be attributed to SSC photoionization alone.
\end{abstract}
\keywords{ galaxies: evolution – galaxies: starburst – methods: numerical – hydrodynamics – ISM: jets and outflows}

\section{Introduction}

Outflows from galaxies and super star clusters (SSC) are a pervasive feature of star forming regions across all redshifts \citep{Heckman1990,Lehnert1996,Heckman2000,Pettini2002,Martin2005,Rupke2005,Veilleux2005,Weiner2009,Bordoloi2014,Bordoloi2016,Rubin2014,Heckman2015,Chisholm2016}.
These outflows are powered by a variety of mechanisms including supernovae \citep{Maclow1999,Scannapieco2001,Mori2002,Scannapieco2002,Springel2003,Dalla2008,Creasey2013}, stellar winds \citep{Hopkins2012,Muratov2015,Hayward2017}, radiation pressure \citep{Thompson2005,Hopkins2011,Murray2011},
cosmic rays \citep{Socrates2008,Uhlig2012,Farber2018},
and hot gas produced by gravitationally-driven motions \citep{Sur2016}.
Outflows have a dramatic influence on their host galaxies, by slowing their chemical evolution \citep{Tremonti2004,Oppenheimer2009,Dave2011,Lu2015,Agertz2015} and either suppressing star formation \citep{Somerville1999,Cole2000,Scannapieco2001,Scannapieco2002,Benson2003} or enhancing star formation \citep{Scannapieco2004b,Gray2010,Gray2011a,Gray2011b,Bieri2016,Fragile2017,Mukherjee2018}.
Radiative and mechanical feedback from stellar clusters is a major driver of galaxy evolution and is especially important in the early universe, particularly during the epoch of reionization \citep[\eg][]{Lehnert2010}.

The nature of these outflows is complex, as the outflow gas is found over a wide range of temperatures.
A complete picture is found only when observations and modeling encompass this full range, from X-ray observations of 10$^{7}$-10$^{8}$ K gas \citep{Martin1999,Strickland2007,Strickland2009}, near-UV and optical observations of $\approx$10$^{4}$ K \citep{Pettini2001,Tremonti2007,Martin2012,Soto2012,Bik2018}, and IR and submm observations of molecular gas at 10-10$^{3}$ K \citep{Walter2002,Sturm2011,Bolatto2013,Leroy2018}.
Most observational studies have focused on the near UV and optical regime due to the strong rest frame emission and absorption lines. 
X-ray observations, on the other hand, can be obtained only for the brightest nearby objects \citep{Lehnert1999,Strickland2009}.

The classic picture of stellar feedback from the parent stellar clusters is that of strong outflows which expel the residual gas, clear pathways for Lyman continuum (LyC) photons to escape their host galaxies \citep{Heckman2011,Zastrow2013} and drive cosmic reionization \citep{Faucher2009,Lehnert2010,Bouwens2012}.
However, there is a growing body of evidence that the outflows are suppressed for the most massive and compact super star clusters \citep[SSCs;][]{Turner2003,Smith2006,Oey2017,Jaskot2017}. 
This may be caused by very dense, overpressurized environments, or the outflows themselves, inducing catastrophic cooling \citep{Wang1995a,Silich2004,Krause2014,Silich2017,Silich2018,Yadav2017}.
The large pressure associated with such high densities, along with contributions from turbulence or gravitational sources, may also suppress these outflows through pressure confinement \citep[\eg][]{Silich2007}.

Several observations now seem to support this suppressed superwind scenario.
\cite{Turner2003,Turner2017} studied an SSC still embedded in a molecular cloud in NGC 5253 and found very narrow molecular line widths that imply velocities far lower than that expected for superwinds from a 10$^{5}$ M$_{\odot}$ cluster.
\cite{Cohen2018} also studied {this cluster and found evidence that in this case the winds from the embedded massive stars may have stalled and a cluster-scale superwind is suppressed.}
Similar molecular kinematics are seen in Mrk~71, an SSC embedded in a dense, 2-pc \HII\ region in NGC 2366 \citep{Oey2017}.
Using optical spectroscopy of several SSCs in M82, \cite{Smith2006} found unusually high thermal and turbulent pressures that may have caused the incipient pressure-driven bubbles to stall \citep{Silich2007}, thereby preventing the launch of superwinds.
Of particular interest, many of the so-called ``Green Pea'' galaxies show low outflow velocities and therefore may represent extreme starbursts with suppressed superwinds \citep{Jaskot2017}.
These galaxies are excellent analogs of high redshift galaxies due to their low metallicities, high UV luminosities, and large specific star formation rates \citep{Cardamone2009,Izotov2011,Amorin2012}.

If catastrophic cooling is responsible for suppressing superwinds from SSCs, then the strong radiation from such cooling may generate observable signatures of this process.  
Knowledge of the emitted line radiation can thus help to discriminate between cooling versus pressure-confinement as the dominant effect in this process.  
With this goal, we present some first-order calculations for a set of one-dimensional, spherically symmetric simulations that study the hydrodynamic, thermal, and ionization state of SSC outflows over a range of outflow velocities, and we present predictions for their line radiation. 

We implement a wind boundary condition in the simulation domain that reproduces outflow properties at the star cluster surface \citep{Chevalier1985,Canto2000,Silich2004}.
The ionization state of the gas is evolved using a non-equilibrium atomic chemistry package that tracks the evolution of several astronomically important atomic ions and includes collisional ionization due to electrons, electron recombination, photoionization, photoheating, and ion-by-ion cooling. 

The structure of the paper is as follows: an overview of galaxy outflows and possible cooling regimes is given in \S2. 
\S3 discusses the model framework and initial conditions along with the hydrodynamic results. 
The line emission is presented in \S4, and the summary and conclusions are given in \S5.  

\section{Outflow Structure and Cooling Regimes}
\begin{figure}
\begin{center}
\includegraphics[trim=0.0mm 0.0mm 0.0mm 0.0mm, clip, width=0.85\columnwidth]{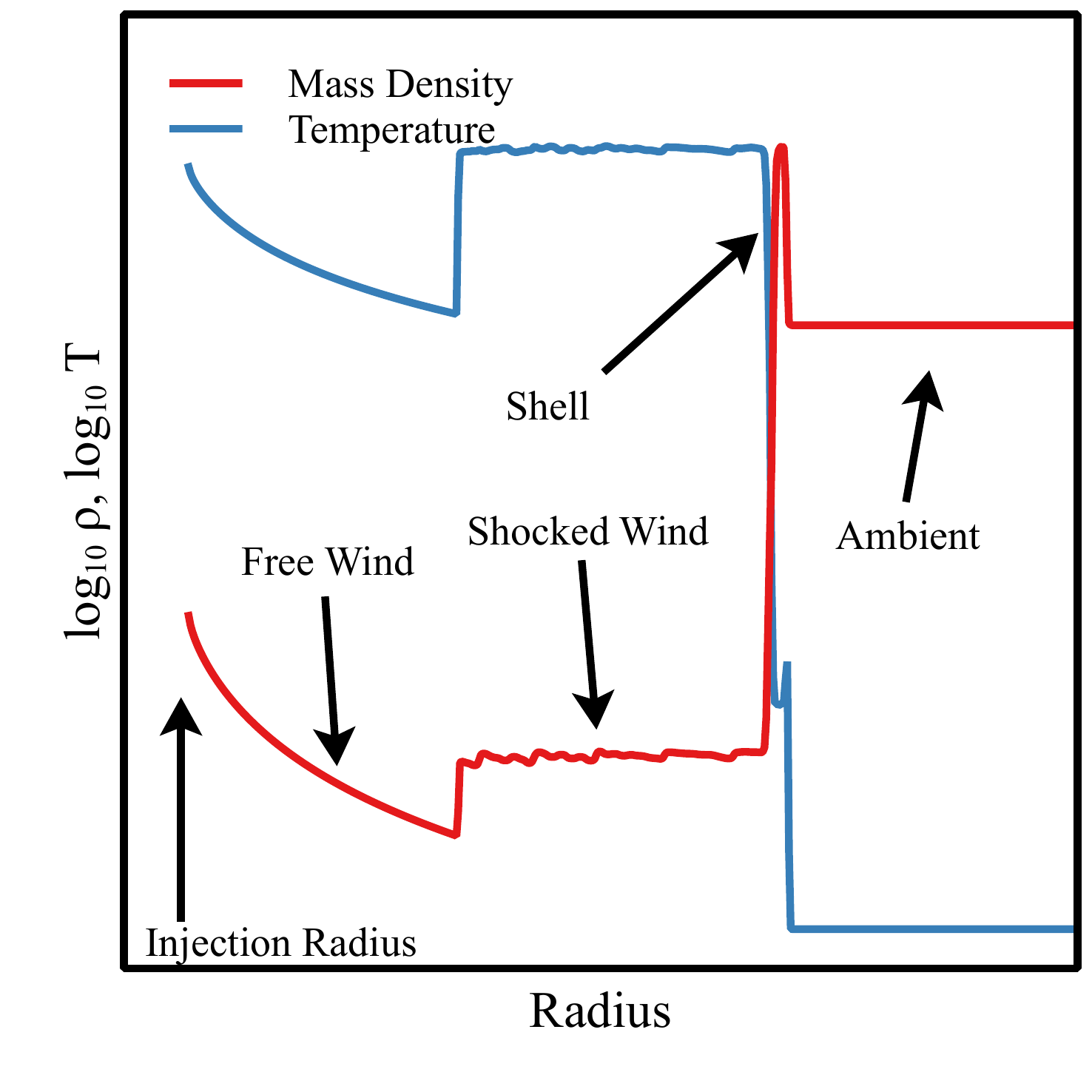}
\caption{Schematic view of the outflow structure, highlighting hydrodynamically important regions. The red line represents the mass density while the blue line represents the gas temperature. Labels give the four typical regions within the outflow, reproducing features described in \cite{Weaver1977}.}
\label{fig:schematic}
\end{center}
\end{figure}

The equations of mass, momentum, and energy conservation in the case of the steady-state outflows are:
\begin{equation}
\frac{1}{r^2}\frac{d}{dr}(\rho u r^2) = q_{m}
\end{equation}
\begin{equation}
\rho u \frac{du}{dr} = -\frac{dP}{dr}-q_{m}u
\end{equation}
\begin{equation}
\frac{1}{r^2}\frac{d}{dr}\left[ \rho u r^2 \left( \frac{1}{2}u^2 + \frac{\gamma}{\gamma-1}\frac{P}{\rho} \right)\right] = q_{e},
\end{equation}
where $r$ is the radial coordinate, $\rho$ is the mass density, $u$ is the velocity, $P$ is the pressure, and $\gamma$ is the adiabatic index.
The mass and energy input rates are 
\begin{equation}
q_{m} = 
\begin{cases}
\dot{M}/V  & \text{if}\ r \le R_{*} \\
0          & \text{if}\ r > R_{*}  \\
\end{cases}
\end{equation}
\begin{equation}
q_{e} = \\
\begin{cases}
\dot{E}/V                 & \text{if}\ r \le R_{*} \\
\sum_i -\Lambda_{i}n_{i}n_{e}  & \text{if}\ r > R_{*}
\end{cases}
\end{equation}
where $\dot{M}$ and $\dot{E}$ are the mass and mechanical energy deposition rates within the cluster, $R_*$ is the sonic radius, $V=4\pi R_{*}^3/3$, $n_i$ is the number density of species $i$, $n_e$ is the electron number density, and $\Lambda_i$ is the cooling function of species $i$.
The classic outflow model of \cite{Chevalier1985}, hereafter CC85, is recovered if one assumes that the outflow persists for many dynamical times and no heating or cooling occurs outside the sonic radius, (\ie\ $q_e$=0, or equivalently $\Lambda_{i}$=0, for $r>R_{*}$). 

Several authors have improved on this model by including heating and cooling within the outflow \citep[\eg][]{Silich2004,Thompson2016}, an incomplete conversion of stellar winds and SN mechanical energy into the outflow energy 
\cite[heating efficiency; \eg][]{Stevens2003,Silich2007,Silich2009,Wunsch2011}, the presence of an ambient medium and radiation pressure effects \citep[\eg][]{Krumholz2009,Martinez2014,Martin2015,Thompson2016,Rahner2017}, the addition of a gravitational potential \citep[\eg][]{Wang1995a,Wang1995b,Bustard2016}, and the effect of non-equilibrium heating and cooling on the ionization state of the outflow \citep{Edgar1986,Shapiro1991,Gnat2007,Kwak2010,Henley2012,Shelton2018,Gray2019}.
For a review of galactic outflows driven by stellar feedback see \cite{Zhang2018}.

The thermalization of stellar winds occurs at reverse shocks between neighboring massive stars in a cluster.
For a $10^{6}\ \rm M_\odot$ SSC with 5 pc radius, the mean half-distance between its $\sim10^4$ massive stars is $X\sim 0.2$ pc.
It takes an extremely short time, a few hundred years ($\tau_{\rm therm}$ $\sim X/v_{\rm inj}$, where the injection velocity $v_{\rm inj}\sim$ 1000 km/s) to thermalize stellar winds.  
A slightly larger time, a few thousand years ($\tau_{\rm hom} \approx 1/3\ R_{\rm inj}/v_{\rm inj}$) is required to fill in the star cluster volume with the gas reinserted by massive stars and form a homogeneous flow.

Figure~\ref{fig:schematic} qualitatively shows the mass density, shown in red, and temperature, shown in blue, of an outflow interacting with an ambient medium where cooling is not dynamically important within the outflow. 
This figure reproduces the classic features found in \cite{Weaver1977}, which are labeled in Figure~\ref{fig:schematic}.
Four distinct regions are formed; the free wind region, the shocked wind, the shock and forward shell, and the undisturbed ambient medium.
The free wind region begins at the sonic radius and is described by its density profile that follows the shape described in CC85.
As the outflow interacts with the ambient medium, a reverse shock is generated that heats the free wind region creating a zone of  nearly uniform density and extreme temperatures. 
In front of this shocked region is a shell of swept up gas from the ambient medium and the forward shock.  
Finally, beyond this shell is the undisturbed ambient medium.

The inclusion of radiative cooling can drastically modify both the temperature and density distribution within the outflow.
Here we are interested in the cooling experienced in the free wind, shocked wind, and within the forward shell and do not consider the flow structure inside the cluster. 
A series of models that aim to study the effect of cooling on the thermal and ionization structure of the outflow and its resulting line emission are presented.

\section{Model Framework and Initial Conditions}
\begin{deluxetable*}{|l|cccccccc|}
	\tablewidth{0.95\textwidth}
	\tablecaption{Simulation Summary}
	\tablenum{1}
	\tablehead{\colhead{Name} & \colhead{$\dot{M}$} & \colhead{$R_{\rm inj}$} & \colhead{$V_{\rm inj}$} & $\eta$ & \colhead{$n_{\rm amb}$} & \colhead{$T_{\rm amb}$}& \colhead{$\rho_{\rm inj}$} & \colhead{T$_{\rm inj}$} \\ 
	\colhead{} & \colhead{(M$_{\odot}$/yr)} & \colhead{(pc)} & \colhead{(km/s)} & \colhead{} & \colhead{(cm$^{-3}$)}  & \colhead{(K)} & \colhead{(10$^{-24}$ gm/cm$^{-3}$)} & \colhead{(10$^{6}$ K)} }
	\startdata
	ETA100    & 10$^{-2}$ & 5 & 1000 & 1.00 & 500 & 10$^{2}$ & 8.4 & 72.6   \\
	ETA025    & 10$^{-2}$ & 5 & 500  & 0.25 & 500 & 10$^{2}$ &16.8 & 18.1   \\
	ETA006    & 10$^{-2}$ & 5 & 250  & 0.06 & 500 & 10$^{2}$ &33.6 & 4.5   \\
	ETA006N1  & 10$^{-2}$ & 5 & 250  & 0.06 & 1   & 10$^{4}$ &33.6 & 4.5   \\
	\enddata
	\tablecomments{ Summary of the simulations presented. The first column gives the name for each model. The second through seventh columns give the mass input rate, injection radius, injection velocity, the heating efficiency, the injection mass density, and density and temperature of the ambient medium respectively. The mass density and  temperature at the injection (sonic) radius is given in columns 8 and 9.}
	\label{tab:runsummary}
\end{deluxetable*}

\begin{figure}
\begin{center}
\includegraphics[trim=0.0mm 0.0mm 0.0mm 0.0mm, clip, width=0.85\columnwidth]{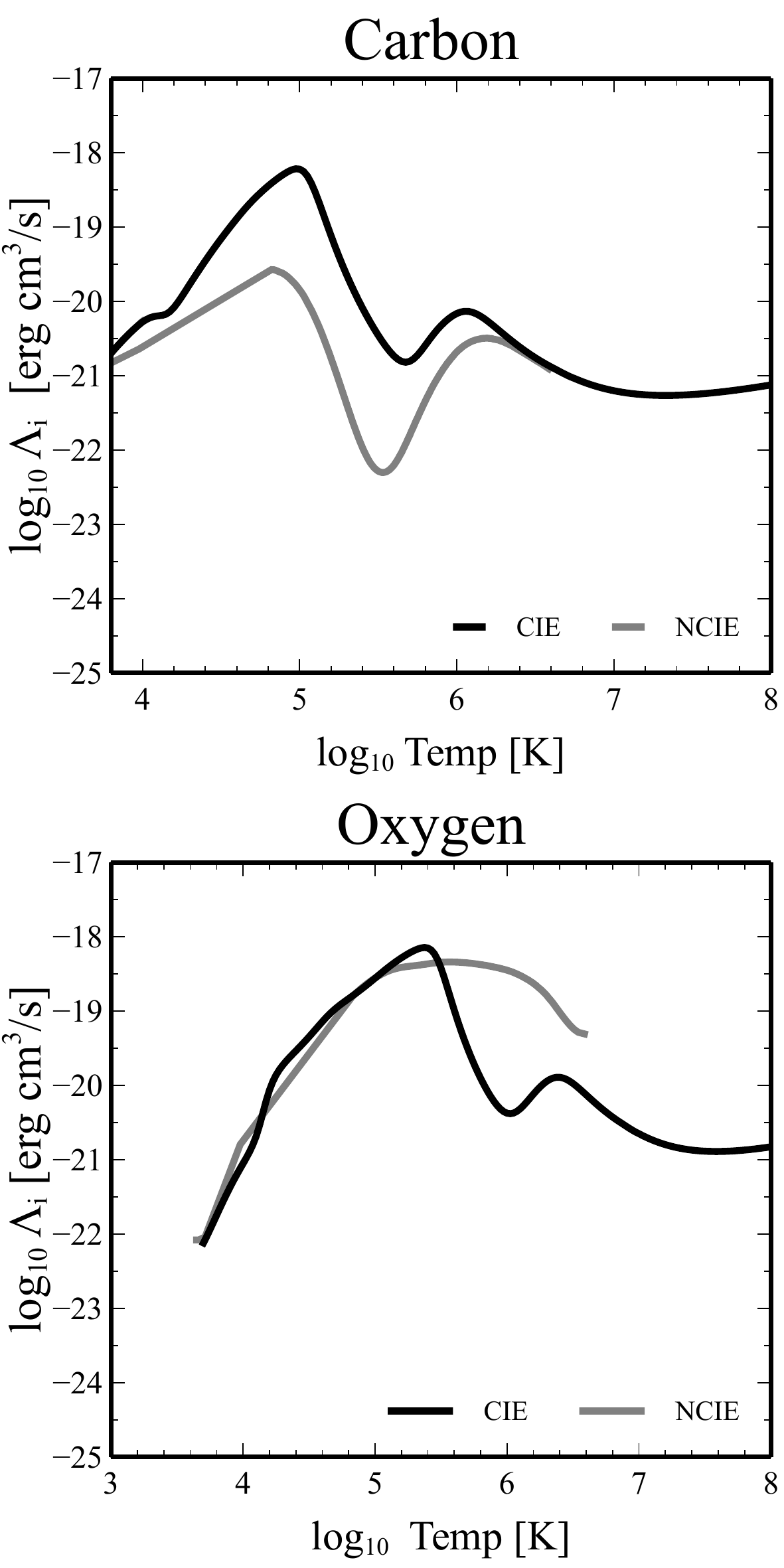}
\caption{ Ion cooling functions for {\it Top Panel:} carbon and {\it Bottom Panel:} oxygen. The black line gives the CIE cooling curve while the gray line shows the non-CIE cooling curve computed using results from ETA006 presented below. Note for carbon the non-CIE rate is substantially lower than the CIE rate. Temperature is given along the $x$-axis while the cooling rate is given along the $y$-axis. }
\label{fig:cooling}
\end{center}
\end{figure}

All of the simulations presented here are performed using \maihem\ \citep{Gray2015,Gray2016,Gray2017,Gray2019} our modified version of the adaptive mesh refinement hydrodynamics code \flash\ \citep{Fryxell2000}. 
One of the primary components of \maihem\ is a non-equilibrium atomic chemistry and cooling package.
This package tracks the evolution of 84 species across 13 atomic elements: hydrogen (\HI-\HII), helium (\HEI-\HEIII), carbon (\CI-\CVII), nitrogen (\NI-\NVIII), oxygen (\OI-\OIX), neon (\NEI-\NEXI), sodium (\NAI-\NAVI), magnesium (\MGI-\MGVI), silicon (\SiI-\SiVI), sulfur (\SI-\SVI), argon (\ARI-\ARVI), calcium (\CAI-\CAVI), iron (\FEI-\FEVI), and electrons (\ELEC). 
For each species we consider collisional ionization by electrons, radiative and dielectronic recombinations, and charge transfer reactions.
Collisional ionization rates are taken from \cite{Voronov1997} while radiative and dielectronic recombination rates are from a series of papers by Badnell and collaborators \citep{Badnell2003,Badnell2006}.
Coupled to this network is a cooling routine that computes the total cooling rate on an ion-by-ion basis.
The procedure used in computing the cooling rates is the same as presented in \cite{Gray2015} and is a reproduction of the results shown in \cite{Gnat2012}.
Examples of the instantaneous cooling rate from ETA006 (presented below) along with the atomic CIE cooling rates are given in Fig~\ref{fig:cooling}.

Radiation pressure from strong UV sources can  affect the hydrodynamics  and ionization state of the outflow. 
For example, it could enhance the gas thermal pressure behind the leading shock at the outer edge of the wind-driven shell \cite[\eg][]{Gonzalez2014,Thompson2015}.
As presented in \cite{Gray2019}, \maihem\ allows the UV background field to vary in space but is assumed to be static in time.
This limitation prevents us from accurately including a UV background in our models and we therefore omit them.
However, as presented below, we can approximate the effect of the UV field on the emission lines using a suite of \cloudy\ models.

\subsection{Boundary Condition}

Classically, the galaxy outflow is described by three parameters, the injection radius $R_{\rm inj}$,  mass input rate {$\dot{M}$}, and the mechanical luminosity $L_{mech}$, of the enclosed cluster.
We also consider two additional parameters, one that describes the density of the ambient medium, $n_{\rm amb}$  and one that describes the initial temperature of the ambient medium $T_{\rm amb}$. 
The energy flux through the surface with radius $R_{inj}$ is:
\begin{equation}
\dot{E} = \eta L_{mech} = \eta\dot{M}\left( v_{\rm inj}^2 /2 + 3 c_{\rm s}^2 /2 \right),
\end{equation}
where $c_s$ is the local sound speed and $\eta$ is the heating efficiency, which determines how efficiently $L_{mech}$ is converted to outflow energy \cite[\eg][]{Wunsch2011}.  
In our models, this is parameterized by varying $v_{\rm inj}$.
The outflow is implemented as an inflow boundary condition, following \cite{Gray2019}.

Following CC85, we assume that the injection radius is coincident with the sonic radius and the injection velocity is, therefore, equal to the local sound speed (i.e, the gas enters the simulation domain with a Mach number of 1).
The injection velocity is related to the enclosed cluster mechanical luminosity as,
\begin{equation}
\label{eqn:vinj}
c_s^2=v_{\rm inj}^2 = \eta\dot{E}/2\dot{M}.
\end{equation}
The mass density of the outflow is defined as
\begin{equation}
\label{eqn:density2}
\rho_{\rm inj} = \dot{M}/\Omega R_{\rm inj}^2v_{\rm inj},
\end{equation}
or conversely,
\begin{equation}
\label{eqn:density}
\dot{M}= \Omega R_{\rm inj}^2 \rho_{\rm inj}  v_{\rm inj},
\end{equation}
where $\Omega R_{\rm inj}^2$ is the effective surface area of the outflow.
For a perfectly isotropic outflow, $\Omega=4\pi.$ Here we take $\Omega=\pi$ following \cite{Gray2019} and \cite{Scannapieco2017}, based on the expectation that superwinds generally break out in preferred directions perpendicular to galaxy disks.

The initial ionization state of the outflow is determined by the initial temperature of the outflow and is given by
\begin{equation}
\label{eqn:temp}
T_{\rm inj} = v_{\rm inj}^2 \bar{A} m_{\rm H} / k_{\rm B},
\end{equation}
where $m_{\rm H}$ is the mass of hydrogen, $\bar{A}$ is the average atomic weight, and $k_{\rm B}$ is Boltzmann's constant. 
The atomic composition and total metallicity of the outflow is assumed to have a default solar value.
Following \cite{Gray2019}, we assume that the incoming gas is in collisional ionization equilibrium (CIE) that depends only on the injection temperature and, when applicable, the UV field.
The initial CIE values are computed using \cloudy\  \citep[e.g.,][]{Ferland2013}.
A table is generated that gives the CIE values as a function of temperature for each of the ionization states tracked by \maihem.
This table is then read by \maihem\ at runtime and linearly interpolated to give the initial ionization state of the inflowing gas. 

We caution that the flow parameterization from the sonic point in terms of $\dot{M}$ and $\eta\dot{E}$ is simplistic and is designed specifically to generate a range of outflow conditions, including catastrophic cooling.  
It is known that strong cooling and other effects also may be important inside the injection radius, which would further suppress the outflow \citep[e.g.,][]{Wunsch2011, Silich2017}.  Therefore, the simulations presented here should be taken fairly qualitatively, as a demonstration of general effects.

\subsection{Simulation Setup}
\label{sec:SimulationSetup}

Each simulation is run in one-dimensional, spherical coordinates with the inner radius equal to the injection radius $R_{\rm inj}$, which is also defined to be the sonic radius and set to 5 pc by default; the outer radius is 75 pc.
Initially, the entire domain is considered to be part of the ambient medium. 
The initial density is set to $n_{\rm amb}$ with an initial temperature of 100 K with initial abundances in their CIE states.
The base grid is comprised of 256 blocks and we allow up to two additional levels of refinement.
The grid is allowed to adaptively refine based on gradients in the density, radial velocity, and gas temperature. 
This gives a maximum resolution of 0.14 pc. 
Each simulation is run for 1 Myr, which is much longer than the typical dynamical time of the outflow, $\tau_{\rm dyn}\approx R_{\rm inj}/v_{\rm inj}$.

As described above, there are five parameters that define each simulation: the injection (sonic) radius; the heating efficiency or, conversely, the injection velocity; the mass input rate; and the density and initial temperature of the ambient medium.
Table~\ref{tab:runsummary} gives a summary of the simulations presented here.
We have chosen a small suite of models that fix the mass input rate at 10$^{-2}$ M$_{\odot}$/yr, the injection radius at 5 pc, and the outflow velocity at 1000 km/s.
We vary the heating efficiency $\eta$ between 1 and 0.06, which corresponds to equivalent injection velocities that range between 1000 km/s and 250 km/s.
We also assume an ambient medium that is initially neutral with a density of 500 cm$^{-3}$ and temperature of 100 K. 
We address the possibility of an ionized ambient medium in Section~\ref{sec:photoionization} below.
Table~\ref{tab:runsummary} also gives the mass density and gas temperature at the injection (sonic) radius as computed by Eqn.~\ref{eqn:density2} and Eqn.~\ref{eqn:temp} respectively.
As we will show below, these parameters create conditions where cooling becomes hydrodynamically and chemically important in the outflow regimes introduced above.  

\subsection{Hydrodynamic Models}
\begin{figure*}
\begin{center}
\includegraphics[trim=0.0mm 0.0mm 0.0mm 0.0mm, clip, width=0.85\textwidth]{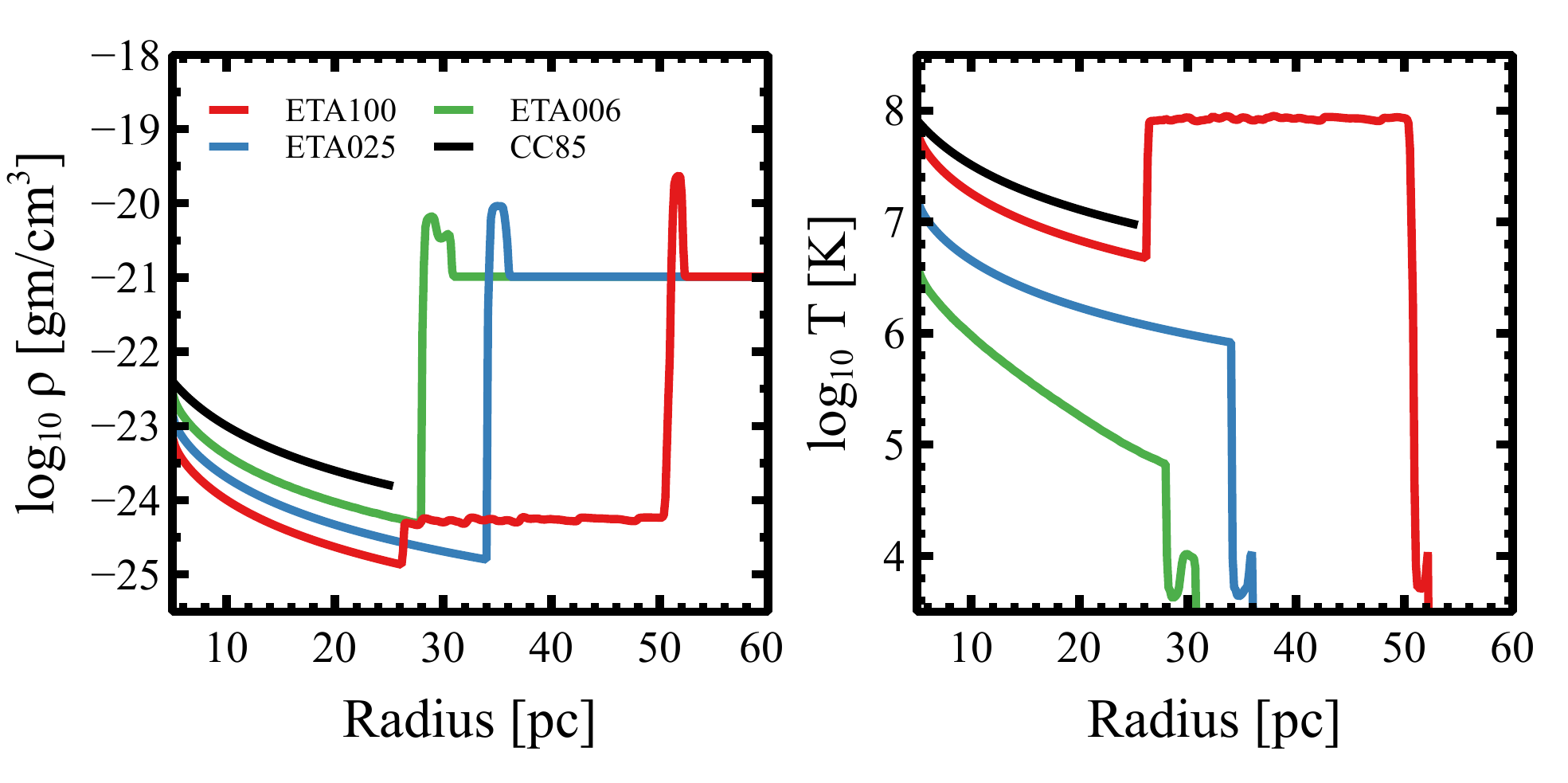}
\caption{ {\it Left Panel:} Logarithmic mass density and {\it Right Panel:} logarithmic temperature profiles for each model at 1 Myr. The legend gives the line color for each model. The black line in each panel represents the density and temperature profiles expected from the \cite{Chevalier1985} adiabatic model solution. }
\label{fig:rhotemp}
\end{center}
\end{figure*}

The mass density and temperature profile for each model are shown in Figure~\ref{fig:rhotemp}. 
The importance of the injection velocity is readily apparent in both panels.
As shown in Table~\ref{tab:runsummary} and by Eqn.~\ref{eqn:density2}, the injection density is inversely proportional to the injection velocity, ranging between 8.4$\times$10$^{-24}$ and 33.6$\times$10$^{-24}$ gm cm$^{-3}$ for the injection velocities considered.
The temperature, however, is quadratically dependent on the injection velocity (Eqn.~\ref{eqn:temp}) creating a much larger range of initial temperatures. 
As we will explore below, this change in initial temperature dramatically changes the initial ionization state of the inflowing gas.

ETA100 reproduces the density and temperature profiles found in \cite{Weaver1977} and CC85. 
This model shows prominent free wind regions with profiles in density and temperature consistent with the adiabatic profiles found in CC85, which is shown by the black line in both panels of Figure~\ref{fig:rhotemp} for comparison. 
This model also reproduces a notable shocked wind region with nearly uniform densities and extremely high temperatures, $>$10$^{8}$ K.

ETA025, however, shows dramatic differences in both the density and temperature when compared to ETA100, and represents a regime in which cooling is important within the shocked wind region.  
Specifically, Figure~\ref{fig:rhotemp} shows that radiative cooling causes the shocked wind region to collapse and it is completely absent in this model.
This produces a density and temperature profile where only the free wind and forward shock / shell regions are present.
Note that the absence of the shocked wind region allows the outflow to expand farther downwind compared to ETA100:
the free wind region extends to R$\approx$32 pc for ETA025, while it only extends to $\approx$27 pc for ETA100.
In this model, cooling within the free wind region is dominated by the expansion of the outflow, since the temperature profile matches that found by assuming adiabatic expansion.

ETA006 represents a transition where cooling within the free wind itself is important.
Figure~\ref{fig:rhotemp} shows that when compared to ETA025, ETA006 is very similar in terms of the density structure and the absence of the shocked wind region.
ETA006, however, shows a steeper temperature profile within the free wind region that indicates that atomic cooling is significant in this region.
The low temperature within the forward shock in all three models is set by our cooling functions which are only defined for $3.7<\log T<8$.  

Figure~\ref{fig:rhotemp} highlights the dramatic effect that simply changing the outflow conditions have on its resulting evolution.
By changing the outflow velocity, the outflow can either produce the classic adiabatic evolution or undergo catastrophic cooling.
The reason for this is twofold.
First, the outflow density is slightly higher for the lower velocity outflows, which increases the cooling efficiency since the cooling rate depends on $n^{2}$.
This represents an increase in cooling efficiency of $\approx$16 between ETA100 and ETA006.
Secondly, for solar metallicity gas, the cooling efficiency is higher at $T\approx$10$^{6}$ K than at $T\approx$10$^{7}$ K by roughly a factor of ten.
This can be seen in the cooling curves of \cite{Sutherland1993} and \cite{Gnat2012}.
Together, this leads to a roughly a 160$\times$ increase in the total cooling rate and the subsequent catastrophic cooling
\citep[\eg][]{Sutherland1993,Gnat2007,Vasiliev2011,Gnat2017}.

\section{Line Emission}
\subsection{Ionization and Emission Due to Cooling}
\label{sec:lineemission}

\begin{figure*}
\begin{center}
\includegraphics[trim=0.0mm 0.0mm 0.0mm 0.0mm, clip, width=0.85\textwidth]{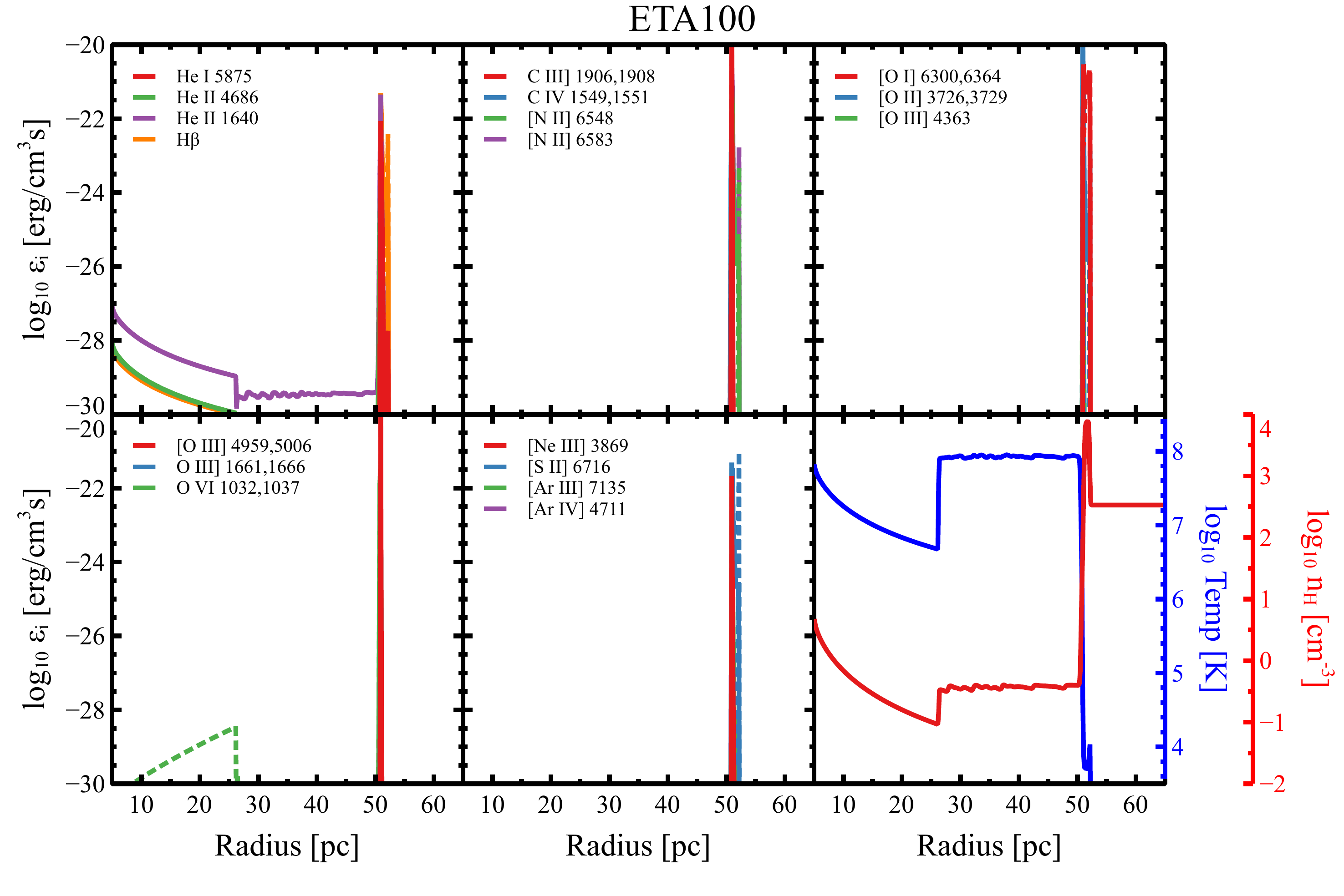}
\caption{ Line emission results for ETA100. The legend in each panel gives the emission lines plotted. The solid lines show the non-CIE \maihem\ results while the dashed lines show CIE results from \cloudy. Here, these results are nearly identical and are coincident with each other. The $x$-axis is the radial distance from the source of the outflow and the $y$-axis gives the logarithm of the line emissivity with units of erg cm$^{-3}$ s$^{-1}$. The  position of the forward shock is at 50 pc. The bottom right panel shows both the temperature density profiles. The blue line shows the temperature, while the red line shows the hydrogen number density. 
}
\label{fig:R5V1000Lines}
\end{center}
\end{figure*}

\begin{figure*}
\begin{center}
\includegraphics[trim=0.0mm 0.0mm 0.0mm 0.0mm, clip, width=0.85\textwidth]{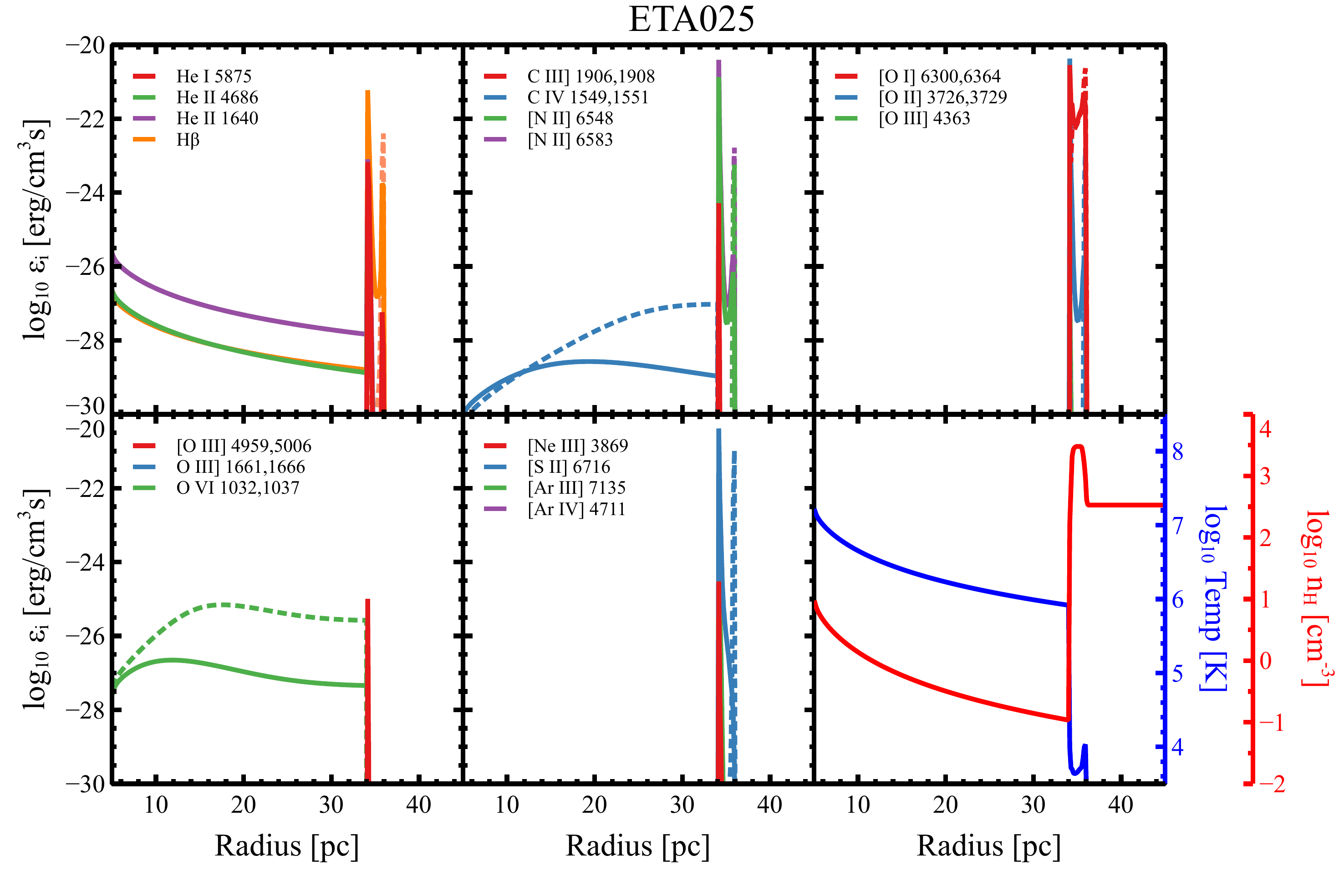}
\caption{ Same as Figure~\ref{fig:R5V1000Lines} for model ETA025. In many cases, the equilibrium and non-equilibrium results are coincident. The final position of the forward shock is found at 37 pc.  }
\label{fig:R5V500Lines}
\end{center}
\end{figure*}

\begin{figure*}
\begin{center}
\includegraphics[trim=0.0mm 0.0mm 0.0mm 0.0mm, clip, width=0.85\textwidth]{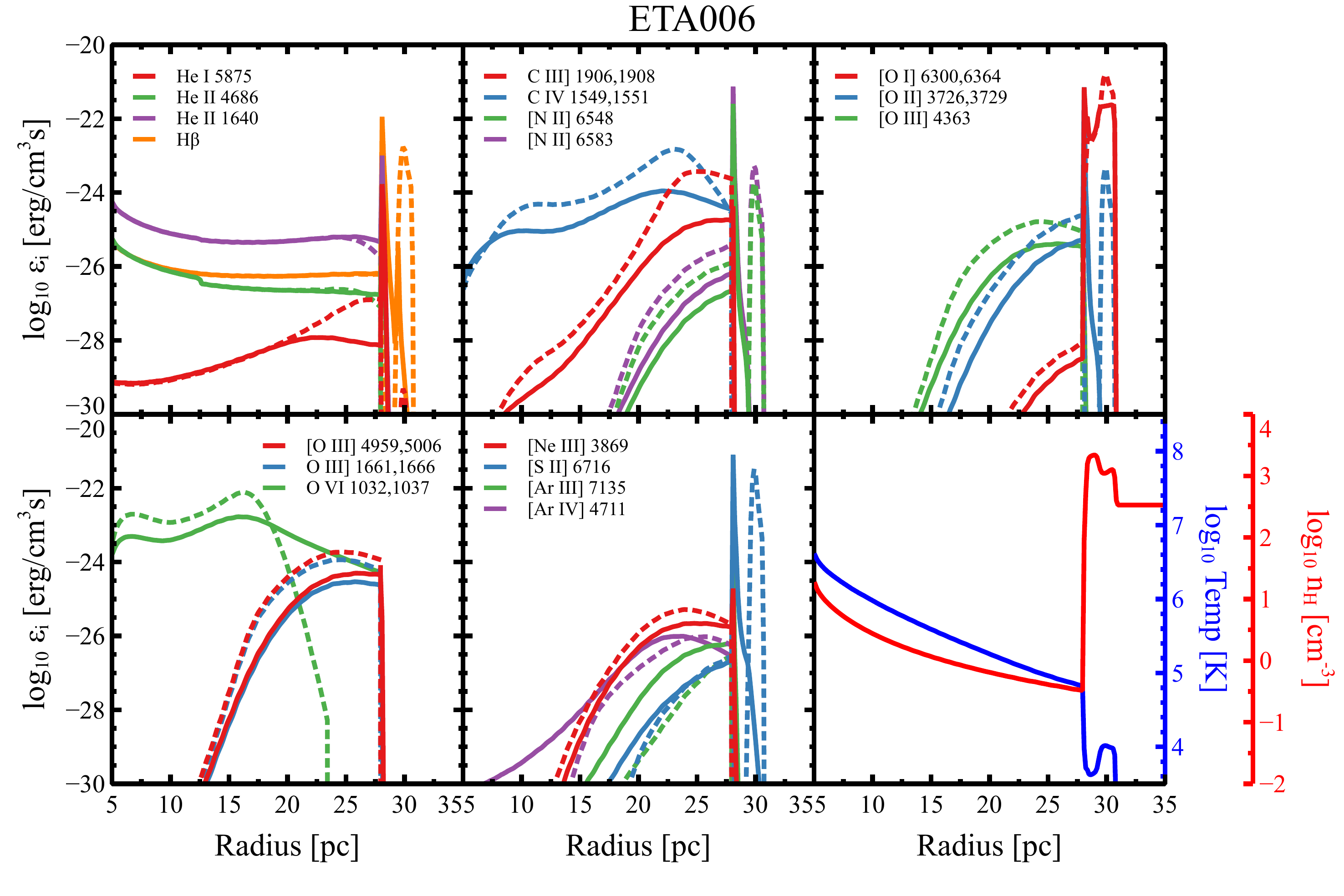}
\caption{ Same as Figure~\ref{fig:R5V1000Lines} for model ETA006.  The final position of the forward shock is found at 28 pc.  }
\label{fig:R5V250Lines}
\end{center}
\end{figure*}

\begin{figure*}
\begin{center}
\includegraphics[trim=0.0mm 0.0mm 0.0mm 0.0mm, clip, width=0.85\textwidth]{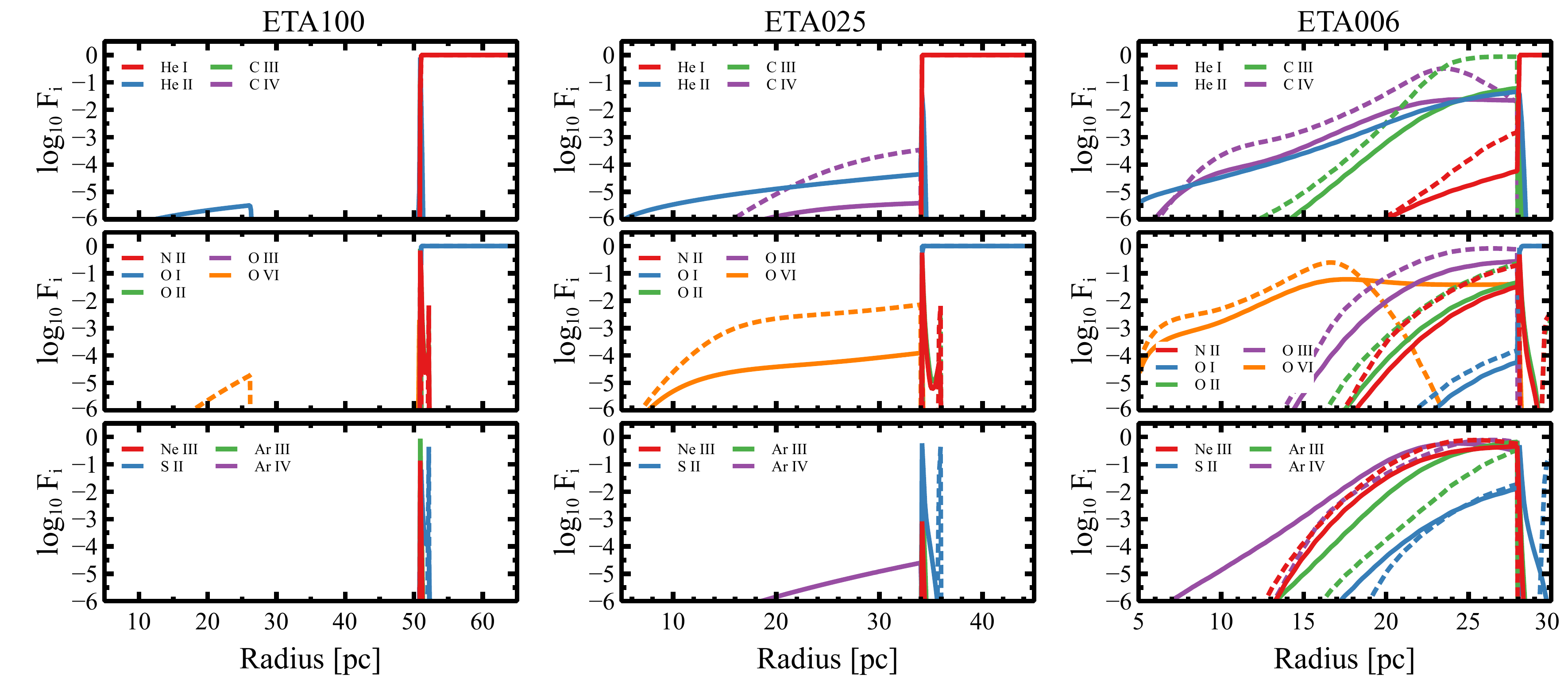}
\caption{ Fractional ionization comparisons as a function of radius for ETA100 (left panel), ETA025 (middle panel), and ETA006 (right panel). The legend in each panel describes the ions. Solid lines show the non-equilibrium results from the \maihem\ models while dashed lines are equilibrium results computed using \cloudy. Note that CIE conditions tend to predict outflows with lower ionization states compared to the non-CIE outflows.}
\label{fig:chemcomparison}
\end{center}
\end{figure*}
\begin{deluxetable}{|l|cc|}
	\tabletypesize{\scriptsize}
	\tablewidth{0.75\textwidth}
	\tablecaption{ Line Luminosities due to Hydrodynamics Only}
	\tablenum{2}
	\tablehead{\colhead{Emission Line} & \colhead{ETA006-CIE} & \colhead{ETA006-nCIE} }
	\startdata
       He I \                  $\lambda$5875                 & -3.35     & -1.78       \\
       He II\                  $\lambda$1640                 & -1.41     & -0.96       \\
       He II\                  $\lambda$4686                 & -2.73     & -1.82       \\
       C III\rbrack\           $\lambda\lambda$1906,1908     &  0.14     & -1.37       \\
       C IV \                  $\lambda\lambda$1549,1551     &  0.59     & -0.54       \\
       \lbrack N II\rbrack\    $\lambda$6548                 & -1.04     &  0.31 \\
       \lbrack N II\rbrack\    $\lambda$6583                 & -0.16     &  1.15 \\
       \lbrack O I\rbrack\     $\lambda\lambda$6300,6364     &  2.13     &  1.49 \\
       \lbrack O II\rbrack\    $\lambda\lambda$3726,3729     & -0.55     &  0.76 \\
       \lbrack O III\rbrack\   $\lambda\lambda$1661,1666     & -0.29     & -1.14       \\
       \lbrack O III\rbrack\   $\lambda$4363                 & -1.14     & -1.99       \\
       \lbrack O III\rbrack\   $\lambda\lambda$4959,5006     & -0.08     & -0.91       \\
       O VI\                   $\lambda\lambda$1032,1037     &  0.84     &  0.29 \\
       \lbrack Ne III\rbrack\  $\lambda$3869                 & -1.15     & -1.61       \\
       Si III                  $\lambda$1206                 & -0.15     & -1.50       \\
       Si IV                   $\lambda\lambda$1393,1403     & -0.11     & -1.09       \\
       \lbrack S II\rbrack\    $\lambda$6716                 &  1.14     &  0.83 \\
       \lbrack S II\rbrack\    $\lambda$6730                 &  0.98     &  0.68  \\
       \lbrack Ar III\rbrack\  $\lambda$7135                 & -3.31     & -2.02       \\
       \lbrack Ar IV\rbrack\   $\lambda$4711                 & -2.40     & -2.64       \\
       \lbrack Ar IV\rbrack\   $\lambda$4740                 & -2.49     & -2.72       \\
       \enddata
	\tablecomments{Emission line luminosities for ETA006-CIE and ETA006-nCIE presented in \S~\ref{sec:lineemission}. Line emission is given as log$_{10}$($\epsilon_i$/$\epsilon_{H\beta}$).  The total H$\beta$ luminosity is 1.8$\times$10$^{36}$ erg/s and 3.3$\times$10$^{36}$ erg/s for ETA006-CIE and ETA006-nCIE respectively. First column gives the particular line while the second and third column gives the results for ETA006-CIE and ETA006-nCIE respectively. Note that emission line luminosities from doublet lines (denoted with $\lambda\lambda$) is computed as the sum of the components. }
	\label{tab:modelLuminosity41}
\end{deluxetable}

The strong cooling described in these models corresponds to radiation that may be characterized by its line emission.
As shown above, the SSC outflows are found to have a wide range of temperatures. 
A complete picture is possible only when the properties of the outflow are measured and modeled over the full range of temperatures.
Observational studies, however, often focus on the warm gas at $\approx$ 10$^{4}$K, where strong emission and absorption lines are found in the rest-frame UV and optical.  This regime is also expected to dominate radiative cooling for strongly cooling flows \citep[e.g.,][]{Gupta2016, Pellegrini2019}.
We have therefore selected a set of UV/optical emission lines to characterize each outflow.  These are listed in Table~\ref{tab:modelLuminosity41}. 

The line emission for each of our simulations is estimated using \cloudy.
By default, \cloudy\ assumes that the ionization state of the gas is in statistical and thermal equilibrium, which, as shown below and in \cite{Gray2019}, may often not be met, especially in strongly cooling outflows.
We therefore present two sets of \cloudy\ line emission results, one in which the ionization states are in CIE, and one with non-equilibrium ionization states.
To run the non-CIE case, the ionization states computed by \maihem\ are input into \cloudy\ through the use of the ``element name ionization'' command, which freezes the ionization state distribution for a given element and requires \cloudy\ to use this distribution when computing the line emission.  It is therefore necessary to perform single-zone \cloudy\ simulations for each of the $\sim$500 radial steps in the \maihem\ model.
In the CIE case, \cloudy\ computes the ionization state of the gas.  To optimize the comparison between these models, we use the same set of radial zones used in the non-CIE model to compute the ionization and emissivities.

Figures~\ref{fig:R5V1000Lines}-\ref{fig:R5V250Lines} show the calculated line emission as a function of radius for each of our models.
The solid lines represent the line emissivities computed using the non-CIE, \maihem\ ionization states and the dotted line shows the equilibrium \cloudy\ results.
The legend in each panel shows the modeled emission lines. 
Appendix~\ref{apdx:LineEmission} shows similar figures but normalized by H${\beta}$.
Figure~\ref{fig:chemcomparison} shows a comparison between the fractional, non-CIE ionization states found in the \maihem\ models and the CIE ionization states computed by \cloudy.

In the case of ETA100, the \maihem, non-CIE and \cloudy, CIE results are essentially identical, which is shown in the left panel of Figure~\ref{fig:chemcomparison}. 
As shown in Figure~\ref{fig:rhotemp}, this model has the lowest outflow density and the highest outflow temperature.
This creates a hot environment with very long recombination timescales, approximated as 
$\tau_{rec}=1/\alpha n_{e}\sim 10^{5}\ n_{e}^{-1}$ years, where $\alpha$ is the recombination coefficient and $n_e$ is the electron number density.
Within the free wind region, $n_e$ is quite small, $\sim$10$^{-1}-1$ cm$^{-3}$, leading to recombination timescales on the order of few $\times 10^{5}-$10$^{6}$ years, which is long compared to the dynamical timescale $\tau_{\rm dyn}\sim R_{\rm shell}/v_{\rm inj}\sim 5\times 10^{4}-2\times 10^{5}$ years for the outflows considered here.

This model corresponds to the conventional, adiabatic bubble model, and so the region interior to the shell is dominated by $10^7-10^8$~K gas (Figure~\ref{fig:rhotemp}).
Thus, the emission throughout the free wind region and shocked wind region is orders of magnitude fainter than the shell emission and generally is not observable. 
Only emission from \HEII\ and H${\beta}$ is present throughout the free wind and the shocked wind regions.  
Both the \HEII\ and H${\beta}$ emission track the density profile for this model, reflecting these two distinct regions.
Only in the high-density shell at the forward shock do the other nebular emission lines become apparent, creating very sharp emission features in, \eg\ [\CIII] and [\OIII].

Figure~\ref{fig:R5V500Lines} and the middle panel of Figure~\ref{fig:chemcomparison} show the results for ETA025.
Similar to ETA100, \HEII\ and H${\beta}$ emission, along with emission from \OVI, dominates throughout the free wind region and all other emission lines appear only at the high-density shell. 
Compared to ETA100, the \HEII\ emission in ETA025 shows only a single, continuous profile corresponding to only the free wind region, again reflecting its density profile.
While otherwise similar to ETA100, the emissivity of \HEII\ and H${\beta}$ is at least an order of magnitude greater in this region, and may be easier to detect. 
Although difficult to see in emission, ETA025 shows some differences between the equilibrium and non-equilibrium chemistry and associated line emission.
For example, \CIV\ $\lambda$1549,1551, \lbrack\CIII\rbrack\ $\lambda$1906 and \CIII\rbrack\ $\lambda$1908, are up to two orders of magnitude weaker in the non-equilibrium models compared to the equilibrium results.

Figure~\ref{fig:R5V250Lines} shows the results for ETA006. 
In addition to being the most interesting model hydrodynamically, it is the coolest model and therefore has the strongest nebular line emission.
Nearly every emission line that we compute is strongly emitting throughout the free wind region.
The line profiles have similar shapes, generally increasing with radius in the outflow and peaking at or near the shell, except for the highest ionization states of oxygen.
The strong emission in this model is due to the density and temperature of the outflow, and roughly traces the gas at $T\sim$10$^{5}$ K. 
The lower outflow temperatures and higher electron densities lead to lower recombination timescales.
This allows the outflow gas to recombine into lower ionization states, increasing the abundances of those ions under consideration.

The right panel of Figure~\ref{fig:chemcomparison} shows the ionization state comparison between \maihem\ and \cloudy\ for ETA006. 
While the non-equilibrium and equilibrium models agree to within roughly an order of magnitude for most ionization states, significant differences are seen.
The CIE model tends to favor lower ionization states for the selected ions and an outflow with more neutral gas.
Similarly, the CIE model slightly over-predicts the line emission compared to the non-equilibrium model.

Overall, Figures~\ref{fig:R5V1000Lines}-\ref{fig:chemcomparison} highlight the importance of following the non-equilibrium atomic chemistry in these systems.
Except for ETA100, which has the lowest outflow density and highest outflow temperature, every model produces ionization states that are out of thermal equilibrium.
In general, the non-equilibrium models produce a more ionized medium. 
This directly impacts the line emission produced, such that the CIE models produce stronger emission for several lines.
For example, \CIV\ $\lambda$1549 and \CIV\ $\lambda$1551 are overproduced in both ETA025 and ETA006, while \lbrack\OIII\rbrack\ $\lambda$5006\ and \lbrack\OIII\rbrack\ $\lambda$4363 are overproduced in ETA006. 

Table~\ref{tab:modelLuminosity41} gives the integrated line luminosities from the CIE and non-CIE models for ETA006.
The integrated line luminosity is given by
\begin{equation}
L_i = 4\pi\int_{R_{\rm inj}}^{R_{HII}} \epsilon_i r^2 dr,
\end{equation}
where $r$ is radial distance and $\epsilon_i$ is the volume emissivity of line $i$. 
We integrate over the outflow region from the injection radius up to the boundary of the ionized region, at the shell.
The simulations in this section do not include "precursor" photoionization outside the shell due to the generated radiation \citep[e.g.,][]{Shull1979}, but this photoionized intensity is small relative to the shell emission.

\subsection{Contribution of Photoionization}
\label{sec:photoionization}
\begin{figure*}
\begin{center}
\includegraphics[trim=0.0mm 0.0mm 0.0mm 0.0mm, clip, width=0.75\textwidth]{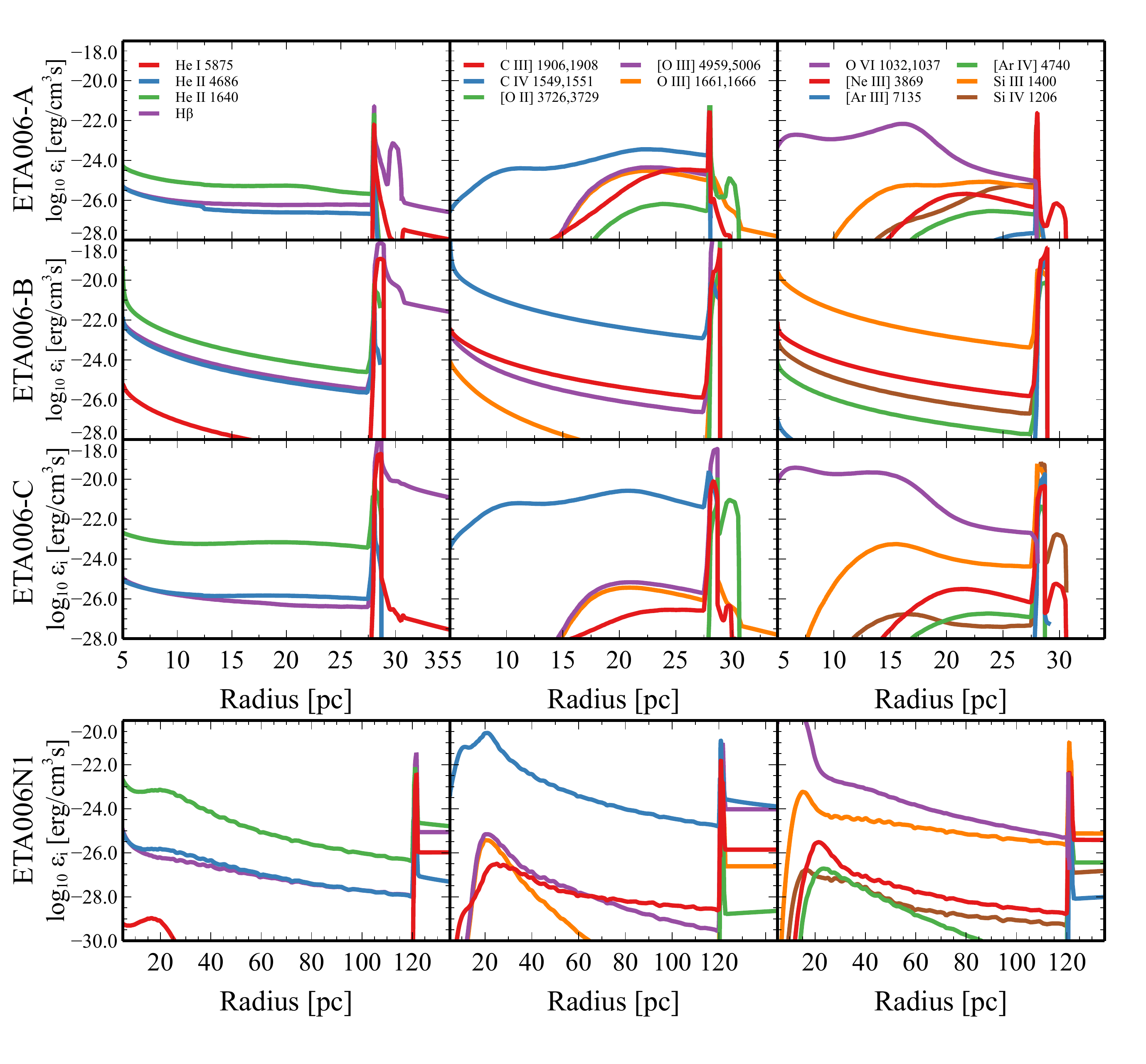}
\caption{ \cloudy\ line emission comparison for ETA006 models. Each column shows a set of emission lines as described by the legend in the first row. The first row shows \maihem\ density and temperature profiles without photoionization (ETA006-A); the second shows \maihem\ density profile only with photoionization (ETA006-B); the third shows \maihem\ density and temperature profiles with photoionization (ETA006-C); and the last row shows the same model based on ETA006 but with an initial ambient temperature of 10$^4$ K and density of 1 cm$^{-3}$ (ETA006N1). The ionized region extends to 2500 pc for ETA006N1. Note that the $x$-axis scale for this model is much larger than for the others. }
\label{fig:R5V250UV}
\end{center}
\end{figure*}

\begin{figure*}
\begin{center}
\includegraphics[trim=0.0mm 0.0mm 0.0mm 0.0mm, clip, width=0.85\textwidth]{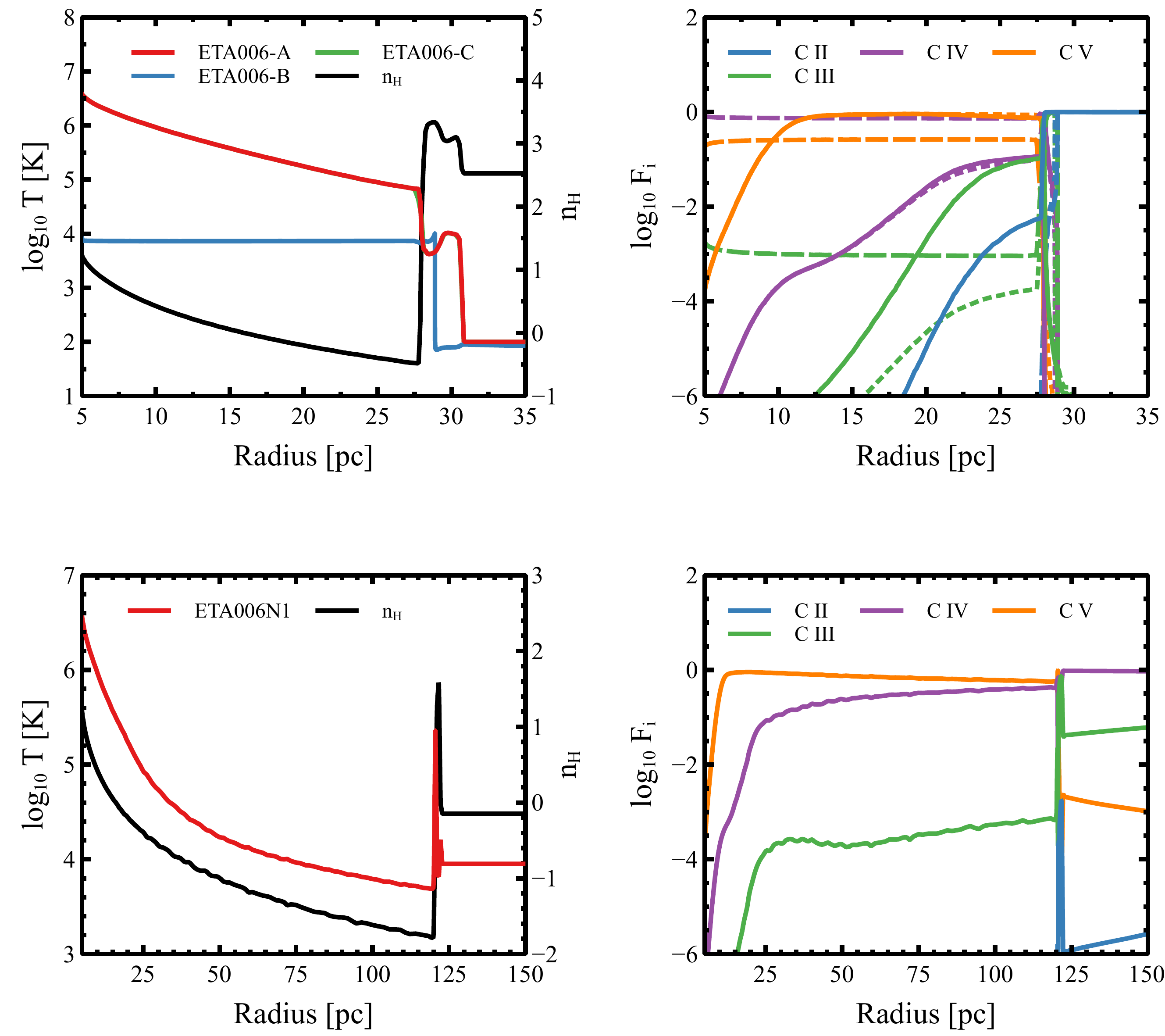}
\caption{ {\it Top Left Panel :} Temperature profiles used as input into the \cloudy\ photoionization models for ETA006 variations.  The solid black line shows the hydrogen number density and is common to each model. Note that ETA006-A and ETA006-C have nearly identical ionization states and therefore are indistinguishable in the figure. {\it Top Right Panel :} Final ionization states for carbon as a function of radius for ETA006-A (solid lines), ETA006-B (dashed lines), and ETA006-C (dotted lines).  Generally ETA006-A and ETA006-C again are nearly indistinguishable. {\it Bottom Left Panel: } Temperature and density profiles used for ETA006N1. {\it Bottom Right Panel: } Final ionization for carbon for ETA006N1C are shown for comparison.   }
\label{fig:CloudyChem}
\end{center}
\end{figure*}

\begin{deluxetable*}{|l|ccc|}
	\tablewidth{0.95\textwidth}
	\tablecaption{Summary of Photoionization Models}
	\tablenum{3}
	\tablehead{\colhead{Name} & \colhead{{\sc MAIHEM} Density} & \colhead{{\sc MAIHEM} Temperature} & \colhead{SB99 UV Field} }
	\startdata
	ETA006-A & \checkmark &  \checkmark &               \\
	ETA006-B & \checkmark &             &  \checkmark   \\
	ETA006-C & \checkmark &  \checkmark &  \checkmark   \\
	ETA006N1 & \checkmark &  \checkmark &  \checkmark   \\
	\enddata
	\tablecomments{Summary of the photoionized models presented. ETA006-A, ETA006-B, and ETA006-C use density and temperature profiles from ETA006. ETA006N1 uses density and temperature profiles taken from a model like ETA006, but with an ambient medium set to 1 cm$^{-3}$. }
	\label{tab:photosummary}
\end{deluxetable*}

\begin{deluxetable*}{|l|ccccc|}
  \tabletypesize{\scriptsize}
  \tablewidth{0.75\textwidth}
  \tablecaption{ Line Luminosities Including Photoionization}
  \tablenum{4}
  \tablehead{\colhead{Emission Line} & \colhead{ETA006-A} & \colhead{ETA006-B} & \colhead{ETA006-C} & \colhead{ETA006N1-R$_{\rm shell}$} & \colhead{ETA006N1-R$_{\rm HII}$} }
  \startdata
    He I \                  $\lambda$5875                   & -1.00    & -0.82    & -0.80    & -1.00    & -0.91 \\
    He II\                  $\lambda$1640                   & -0.30    & -2.54    & -2.45    & -0.60    & -2.41 \\
    He II\                  $\lambda$4686                   & -1.19    & -4.07    & -4.01    & -1.82    & -4.06 \\
    C III\rbrack\           $\lambda\lambda$1906,1908       & -0.04    & -0.95    & -2.20    & -0.50    & -0.58 \\
    C IV \                  $\lambda\lambda$1549,1551       & -0.13    & -2.07    & -1.42    &  0.77    & -1.21 \\
    \lbrack N II\rbrack\    $\lambda$6548                   &  0.07    & -1.63    & -1.71    & -3.50    & -1.37 \\
    \lbrack N II\rbrack\    $\lambda$6583                   &  1.07    & -0.80    & -2.32    & -2.35    & -0.39 \\
    \lbrack O I\rbrack\     $\lambda\lambda$6300,6364       &  1.46    & -2.37    & -1.38    & -7.90    & -3.13 \\
    \lbrack O II\rbrack\   $\lambda\lambda$3726,3729        &  0.69    & -1.01    & -2.64    & -2.74    & -0.78 \\
    \lbrack O III\rbrack\   $\lambda\lambda$1661,1666       & -0.80    & -1.94    & -3.67    & -0.87    & -1.51 \\
    \lbrack O III\rbrack\   $\lambda$4363                   & -1.38    & -1.75    & -3.50    & -1.45    & -1.40 \\
    \lbrack O III\rbrack\   $\lambda\lambda$4959,5006       &  0.29    &  0.88    & -0.52    &  0.68    &  1.07 \\
    O VI\                   $\lambda\lambda$1032,1037       &  0.39    & -9.30    & -1.39    &  0.72    & -1.55 \\
    \lbrack Ne III\rbrack\  $\lambda$3869                   & -0.24    & -0.61    & -2.34    & -0.71    & -0.37 \\
            Si III           $\lambda$1206                  & -0.73    & -1.09    & -1.17    &  0.09    & -1.12 \\
            Si IV           $\lambda\lambda$1393,1403       & -1.36    & -1.30    & -1.37    &  0.36    & -1.24 \\
    \lbrack S II\rbrack\    $\lambda$6716                   &  1.62    & -1.78    & -1.19    & -3.82    & -1.34 \\
    \lbrack S II\rbrack\    $\lambda$6730                   &  1.46    & -1.64    & -1.32    & -3.96    & -1.51 \\
    \lbrack Ar III\rbrack\  $\lambda$7135                   & -0.70    & -1.00    & -1.99    & -1.81    & -0.85 \\
    \lbrack Ar IV\rbrack\   $\lambda$4711                   & -2.84    & -1.88    & -3.25    & -1.62    & -1.66 \\
    \lbrack Ar IV\rbrack\   $\lambda$4740                   & -2.94    & -2.02    & -3.38    & -1.76    & -1.81 \\
  \enddata
  \tablecomments{Line luminosity for \cloudy\ models presented in \S~\ref{sec:photoionization}. Line emission is given as log$_{10}$($\epsilon_i$/$\epsilon_{H\beta}$). ETA006-A has a total H$\beta$ luminosity 5.4$\times$10$^{36}$ erg/s, ETA006-B of 9.5$\times$10$^{40}$ erg/s, and ETA006-C of 9.5$\times$10$^{40}$ erg/s. Two sets of results are shown for ETA006N1, column 5 shows the results where we integrate up through the forward shell and column 6 where we integrate up through the entire \HII\ region. The total H$\beta$ luminosity is 7.8$\times$10$^{38}$ erg/s and 1.5$\times$10$^{41}$ erg/s for each case respectively. Note that emission line luminosities from doublet lines (denoted with $\lambda\lambda$) is computed as the sum of the components. }
  \label{tab:modelLuminosity42}
\end{deluxetable*}
We now aim to estimate the effect of photoionization on the line emission from these outflow models. 
In particular, we evaluate the effect of photoionization using the temperature and density profiles generated by \maihem.
As mentioned above, strong radiation can have important effects on the hydrodynamics of the outflow.  However these effects are absent in our \maihem\ models, as they are unable to carry out the frequency-dependent radiative transfer calculation necessary to handle them properly.

We note that the wind-driven shell never expands beyond the photoionized region in any of the models with the high ambient density \citep[e.g.,][]{Dove2000}, which we examine in \S~\ref{sec:lineemission}. 
In the ETA100 case, the central cluster photoionizes only a small mass fraction of the swept-up shell ($M_{\rm HII}/M_{\rm sh}\sim$0.03) whereas in models ETA025 and ETA006 the ionized gas fraction increases due to smaller wind-driven shell radii and ionized gas densities that drop with $v_{\rm inj}$: $M_{\rm HII}/M_{\rm sh}\sim$ 0.3 and $M_{\rm HII}/M_{\rm sh}\sim$ 0.75 in models ETA025 and ETA006, respectively (see Appendix~\ref{apdx:IonizingRadiation}). 
Nevertheless, in all cases Lyman continuum photons are trapped within the wind-driven shell. 
Note also that photoionization makes the shells thicker, since in the photoionized region the gas temperature is $\sim$ 2.5 times larger than the minimum values obtained in our calculations, resulting in $\sim$ 2.5 times smaller gas density than the maximum values (see Fig.~\ref{fig:rhotemp}).

To estimate the effect of photoionization on the line emission, we post-process the \maihem\ results using \cloudy, now including photoionization from a radiation field representing that of the parent SSC.  
In the previous Section~\ref{sec:lineemission}, the line emission was independently modeled at the individual radial steps, which was necessary to estimate the line emission using the non-equilibrium ionization states produced by \maihem.
Therefore, the line emission for both the CIE and non-CIE models were calculated piecemeal.
However, in this section, modeling the photoionization requires a single, multi-zone \cloudy\ model in order to properly propagate the photoionization outward from $R_{\rm inj}$. 
This also requires that the ionization states be in equilibrium given the imposed radiation field.  
Therefore, in this section, only the radial density and/or temperature profiles from \maihem\ are taken as input, and not the non-equilibrium ionic abundances shown in the previous section.  
The variation between equilibrium and non-equilibrium emission for the photoionized cases can be qualitatively evaluated based on results from Section~\ref{sec:lineemission}. 

We employ Starburst99 \citep[SB99;][]{Leitherer1999} to generate the ionizing spectral energy distribution.
The SB99 model is run with fixed stellar mass of 4.1$\times$10$^{6}$ M$_{\odot}$, solar metallicity, a Salpeter initial mass function, and an age of 1~Myr, consistent with the final age of the \maihem\ simulations. 
This combination of parameters gives a model with a mass loss rate of 10$^{-2}$ M$_{\odot}$/yr and a total luminosity of 10$^{43}$ ergs/s, matching the mass input rate of the \maihem\ models.

As shown above, models ETA100 and ETA025 have temperatures and ionization states that are too high to produce nebular collisional line emission within the free wind region {and hot shocked wind region. 
Thus, the inclusion of photoionization also has no effect on the resulting line emission for these species.
On the other hand, ETA006, with its higher densities, lower temperatures, and lower average ionization states, does produce a variety of line emission throughout the free wind region. 
Therefore, we focus on ETA006 to examine how photoionization affects its line emission.

We compare four variations of this model as follows:
\begin{itemize}
	\item ETA006-A: The hydrodynamic case with no photoionization, but density and temperature profiles taken from \maihem. This model is similar to the CIE model as presented in \S~\ref{sec:lineemission}. We discuss the differences between these models below.
	\item ETA006-B:  A model that includes photoionization, adopting only the density profile from \maihem, and not the temperature profile.  All species are assumed initially neutral and any changes in ionization state are solely due to the UV source.  This model represents the pure photoionization limit.
	\item ETA006-C:  A photoionization model using both density and temperature profiles taken from the \maihem\ model. 
	\item ETA006N1:  The same as ETA006-C but for a model with an initial ambient temperature of $10^4$ K and ambient density of 1 cm$^{-3}$.
\end{itemize}
These models are summarized in Table~\ref{tab:photosummary}.

Figure~\ref{fig:R5V250UV} shows the line emissivities for each of these models in rows 1 -- 4, respectively.
We have chosen to plot a subset of emission lines in order to simplify the presentation.
A version of Figure~\ref{fig:R5V250UV} but with each line normalized by H${\beta}$ is given in Appendix~\ref{apdx:LineEmission}.  
Table~\ref{tab:modelLuminosity42} presents the integrated line luminosities for these models, calculated as before.  
For all models except ETA006N1, the ionized boundary is coincident with the forward shell since the shell is optically thick to LyC photons, as noted above.
For ETA006N1, the shell is optically thin. Column 5 of Table~\ref{tab:modelLuminosity42} gives the luminosities integrated only within the shell outer radius; while column 6 gives values integrated through the entire \HII\ region, which extends to a radius of $\sim$2500 pc.

Model ETA006-A represents the model for emission due only to the hydrodynamics.  
It is the same as the CIE model presented in Section~\ref{sec:lineemission}, but as noted above, it is run as a single \cloudy\ model.
The photoionizing radiation field that \cloudy\ propagates is comprised of two components, an incident field defined by the user, and a diffuse field that is generated by hot gas within the \cloudy\ model.  For ETA006-A, the user-defined field is not included, and therefore it is the diffuse field that generates the minor differences between single-zone (Figure~\ref{fig:R5V250UV}) and multi-zone  (Figure~\ref{fig:R5V250Lines}) emissivities.  
In particular, we see that the ambient medium is now photoionized, as seen in precursor \HII\ regions for shock emission \citep[e.g.,][]{Shull1979}.  The effect of shock expansion into an ionized medium is explored with other models below.

Model ETA006-B is the limiting model for pure photoionization, which we include to show the effects of only photoionization and photoheating (upper-left panel of Figure~\ref{fig:CloudyChem}). 
Whereas the ionization states for model ETA006-A are set by shock-heated temperatures that are 1 -- 2 orders of magnitude greater, the ionization structure of ETA006-B is largely uniform with radius, with the gas dominated by \CIV\ $\lambda$1551 up to the high-density shell (Figure~\ref{fig:R5V250UV}).  
As noted in the previous section, the line emissivities in catastrophically cooling outflows differ markedly from their density profiles due to the temperature distribution of $10^5$~K gas.  
Therefore, the ionization structure of ETA006-B is very different from ETA006-A, as shown for carbon in the upper-right panel of Figure~\ref{fig:CloudyChem} (see also right panel of Figure~\ref{fig:chemcomparison}).

Model ETA006-C combines both the hydrodynamic heating and SSC photoionization.
Comparing ETA006-A and ETA006-C elucidates the contribution of fluorescence and photoionization to the line emission, demonstrating that these significantly boost the line emission of some species.  
For example, \CIV\ $\lambda$1549,1551  and \OVI\ $\lambda1037$ are enhanced by over two orders of magnitude within the free wind region, largely at the expense of \CIII] and [\OIII] (Figure~\ref{fig:R5V250UV}). 
This is true even though the abundance of \CIV\ and the gas temperatures are essentially the same between the two models (Figure~\ref{fig:CloudyChem}, top right panel).  

The integrated line luminosities, however, are dominated by the high-density shell (Table~\ref{tab:modelLuminosity42}).  Because the photoionizing radiation is diluted at the large shell radius and the shell gas density is high, the ionization parameter in the shell is low, resulting in weak [\OIII] and [\ARIV].  Nevertheless, the strong \CIV\ and \OVI\ emission generated by the strongly cooling wind (e.g., Model ETA006-A) maintains line strengths in these ions well above nebular photoionized values.  This suggests that these lines may offer important diagnostics of catastrophically cooling outflows.

A weak, cooling outflow likely expands into an ambient medium that is already photoionized by the SSC.  
We approximate this condition by also running the \maihem\ model for ETA006 but setting an ambient medium temperature of $10^4$~K instead of $10^2$~K.  Given the high ambient density of 500 cm$^{-3}$, this model strongly recombines and thus we find no significant difference in line emissivities between this model and ETA006-C.

Since more realistic ambient densities will strongly decrease as a function of radius from the SSC, we also consider another model, ETA006N1, where the ambient density is lowered to 1 cm$^{-3}$ and the ambient temperature is set to 10$^{4}$ K, but is otherwise identical to ETA006.
We have post-processed ETA006N1 in the same manner as ETA006-C, that is, with the density and temperature profiles generated by ETA006N1 and with the SB99 UV field. 

The results for this model are shown as the fourth row in Figure~\ref{fig:R5V250UV}.
This model is more closely related to ETA025 (Section~\ref{sec:lineemission}), as seen in the temperature and density profile for ETA006N1 shown in the bottom row of Figure~\ref{fig:CloudyChem}.
Unsurprisingly, the outflow expands much farther in ETA006N1 owing to the lower ambient density. 
We also see that the ionization structure now more closely resembles that of the pure photoionization model ETA006-B, particularly at large radii, where the line emissivities generally decrease with radius in the free wind region.
The most prominent difference between ETA006N1 and ETA006-B is strong \OVI\ emission throughout, and the enhanced \CIV\ emission at small radii.  
There is also a central gap in emission for most species.  
ETA006N1 also continues the trend of large \CIV/\CIII\ ratios throughout the free-wind region.

However, in particular, we see that the ambient medium is fully photoionized by the SSC in this model.  
The photoionization extends to a radius of 2500 pc, and so the dense shell would likely appear as a density-bounded \HII\ region.  
In contrast, for ETA006-B, the only low-level photoionization outside the shell is due to the precursor radiation described above. 
Similarly, we showed earlier that ETA025 would have an optically thick shell if photoionized by the SSC.

As noted above, column 5 of Table~\ref{tab:modelLuminosity42} provides total line luminosities for this model integrated only through the outer boundary of the shell; these simulate observations of only the density-bounded, shell \HII\ region, corresponding to spatially resolved observations or data dominated by the high surface-brightness emission.
We see that density-bounded conditions strongly affect emission in \HEII, \CIII], \SiIV, and \SiIII, which are all greatly enhanced.
Column~6 gives line emission integrated through the total, 2500-pc \HII\ region.  While the integrated nebular emission differs significantly depending on the limiting radius, we see that for both tabulated cases, the \CIV\ and \OVI\ emission remain significantly elevated above the pure photoionized values of ETA006-B.

\section{Discussion}

\subsection{Optical Depth and LyC Escape}

We find that in our one-dimensional simulations, the swept-up shell is optically thick to the radiation from the SSC for the default parameters in model ETA006 \citep[cf.][Appendix~\ref{apdx:IonizingRadiation}]{Dove2000}. 
This prevents the cluster radiation from penetrating into the undisturbed ambient medium, which is at odds with observations linking such systems to Green Peas and other LyC emitters \citep{Jaskot2013,Jaskot2017}. 
However, this optical thickness is most likely due to the constrained geometry of 1-D simulations, which does not allow for the possibility of the shell breaking up into clumps. 
Clumping would naturally lead to increased leakage of LyC photons \citep{Jaskot2019}.
In multidimensional simulations, the shell is likely to be susceptible to the Rayleigh-Taylor instability \citep{Rayleigh1883,Taylor1950}, which can break up the shell into clumps on the order of the shell thickness.  
Moreover, catastrophic cooling conditions should also strongly induce the cooling instability, which can increase density perturbations by over an order of magnitude \citep{Scannapieco2017}. 
Thermal instability can cause clumping in the free wind itself \citep[e.g.,][]{Schneider2018, Thompson2016}, or even at radii $< R_{\rm inj}$, thereby decreasing the initial heating efficiency \citep{Wunsch2011}.
However, even when these additional hydrodynamic effects are taken into account, it is possible to generate lines of sight without clumps, or lines of sight with average densities similar to those modeled here.
In this case, we expect the results from these one-dimensional simulations to model these lines of sight.

Evidence does suggest that extreme Green Peas are strongly clumped, consistent with this catastrophic cooling scenario.  A number of studies point to a "picket fence" geometry in these and similar candidate LyC emitters, noted from absorption-line and neutral gas studies \citep[e.g.,][]{Heckman2011,RiveraThorsen2017,McKinney2019}.
\citet{Jaskot2019} show that the SSC environment in extreme Green Peas is consistent with a two-component model consisting of optically thick, high-density clumps at close quarters to the SSCs, which generate the extreme ionization parameters; and an optically thin, interclump medium, through which LyC radiation can escape.

In addition, we do find in some simulations that if the shell expands into a lower density ambient medium, it becomes possible to form an optically thin shell (e.g., ETA006N1; Figure~\ref{fig:R5V250UV}).  
Thus, the optical depths in the simulations presented in this work are not incompatible with suppressed superwinds being linked to LyC emitters.
Future studies will examine these issues using a wider range of outflow parameters.

\subsection{Line Diagnostics of Catastrophic Cooling}

Our models predict that catastrophically cooling winds generate strong \CIV\ and \OVI\ emission that is not observed in ordinary, photoionized \HII\ regions.  \CIII] is strongly elevated in models for cooling outflows expanding into low-density ambient medium, but greatly suppressed when the ambient density is high.
Interestingly, emission in all these lines appears to be linked to the most intense starbursts.  Strong, nebular \CIV, \CIII], and \HEII\ $\lambda1640$ have been reported in, e.g., nearby ($z\lesssim 0.2$) high-excitation, compact dwarf starbursts identified by \citet{Senchyna2017} and \citet{Berg2019a} from SDSS; and also in $z\sim 3$, luminous Ly$\alpha$ emitters selected by \citet{Amorin2017} to have strong \CIII] and \OIII].  

\citet{Hayes2016} present imaging in \OVI\ $\lambda 1037$ of an extreme, nearby starburst, J1156+5008.  While the scale of this starburst and resultant emission is much larger than our models, the qualitative features may be consistent with our predictions.  There is a clear, 9-kpc shell around the starburst.  In one analysis, a central, 1-kpc region centered on the starburst also shows positive emission.  However, the authors suggest that this apparent emission is spurious, based on their lack of confidence in the continuum subtraction method.  We note however, that our models do predict strong central \OVI\ emission, in addition to the bright shell (Figure~\ref{fig:R5V250UV}), and perhaps some nuclear emission is not ruled out in this object (M. Hayes, private communication).

Intriguingly, \citep{Berg2019b} recently reported observations of resonantly scattered \CIV\ line profiles in two local starbursts having very strong \CIV\ $\lambda\lambda 1548, 1550$ emission and \HEII\ $\lambda1640$ emission.  Such resonant scattering implies large column densities of this relatively hot, highly ionized gas, with optical depths of $\tau_{\rm C IV}\sim 10,000$.  Since our models show that \CIV\ is especially prevalent in outflows with strong cooling, rather than conventional, adiabatic superbubble feedback (Figure~\ref{fig:chemcomparison}), could these observations indicate catastrophic cooling?
For a \CIV\ scattering cross-section of $3\times 10^{-14}\ \rm cm^2$ at $\sim 10^5$~K \citep[e.g.,][]{Sankrit2001}, the implied value of $\tau_{\rm CIV}$ requires a \CIV\ density of $n_{\rm C IV}\sim 10^{-2}$ in a 10~pc column.  For C/H abundance on the order of $10^{-4}$, this implies electron densities of $n_e \sim 100 -1000\ \rm cm^{-3}$, depending on the ionic fraction of \CIV.  However, 
\CIV\ does not dominate in the higher density regions.  In our models, high densities are only found at the hot center near the ionizing SSC, and in the cooler shell; the \CIV\ abundance is low in both of these regions (Figure~\ref{fig:CloudyChem}).  

Indeed, high $n_{\rm CIV}$ is extremely difficult to maintain because of the efficiency of the cooling functions at temperatures where \CIV\ dominates (e.g., Figure~\ref{fig:cooling}), explaining the lack of resonantly scattered \CIV\ observations in diffuse gas.  The $\tau_{\rm CIV}$ in our catastrophic cooling models are still generally 2 -- 3 orders of magnitude too small to cause signficant resonant scattering.  Thus, given the difficulty of generating high enough $n_{\rm CIV}$ in starbursts, it may be more likely that the \citet{Berg2019b} observations do not show resonant scattering, and instead show more conventional, kinematic effects.  The line splitting and broader profiles may be due to, e.g., hot, turbulent, bipolar flows.  \citet{Berg2019b} compare the \CIV\ emission-line profiles to those for \OIII], which exists at 10$\times$ lower temperatures and is not strongly co-spatial with \CIV.
On the other hand, there may be extreme catastrophic cooling conditions where $\tau_{\rm CIV}$ is high enough to generate resonant scattering.  Further study of a wider parameter space is needed to evaluate this possibility.

If catastrophic cooling is present in extreme, compact starbursts, then the emission-line spectra will not reflect pure photoionization as is ordinarily assumed when interpreting such spectra.  In particular, the outflow kinematics generate higher ionization states that elevate emission in the corresponding species, as already suggested by, e.g., \citet{Gray2017}.  Thus it may not be necessary to invoke hotter or composite photoionizing sources in extreme starbursts, as is often suggested to be necessary \citep[e.g.,][]{Senchyna2017,Berg2019b,Nakajima2018,Jaskot2013}.  Density-bounded conditions also can drastically affect the line ratios, as seen for the integrated line emission of model ETA006N1 in Table~\ref{tab:modelLuminosity42}.

\section{Summary and Conclusion}

We have presented here a set of one-dimensional models that aim to study a range of radiative cooling regimes within SSC outflows. 
Our models track the non-equilibrium evolution of several atomic species and compute the cooling from these species on an ion-by-ion basis.
The outflow is based on the classic outflow model presented by \cite{Chevalier1985} and \cite{Weaver1977}, and is defined by the mass input rate ($\dot{M}$), the mechanical luminosity ($L_{mech}$), the injection (sonic) radius ($R_{\rm inj}$), the ambient density ($n_{\rm amb}$), and ambient temperature ($T_{\rm amb}$).
Our model defines the energy input rate as a function of the injection velocity.
The outflow model is defined as a boundary condition at $R_{\rm inj}$ of each model, with initial ionization states set to their CIE values.

We present results from four models with the $\dot{M}=10^{-2}$ M$_{\odot}/{\rm yr}$, an injection (sonic) radius of $R_{\rm inj}$=5 pc, an ambient density of $n_{\rm amb}=500$ cm$^{-3}$, and varying the heating efficiency between 1 and 0.06, parameterized by an injection velocity between 1000 and 250 km/s.
With these initial conditions the outflow densities vary by a factor of four while the outflow temperature varies by a factor of sixteen.
This range of energy injection rates generates a wide range of outflow structures.
The classic adiabatic solution corresponds to model ETA100, which has the fastest outflow velocity. 
This model reproduces the three primary outflow features, the free wind region, the shocked wind region, and the swept up shell and forward shock.
For models with slower outflow velocities, ETA025 and ETA006, the shocked wind region fails to form, and for the latter, the free wind region itself is strongly cooling and non-adiabatic.
In these models, it is the properties of the outflow that cause suppressed superwinds by catastrophic cooling, and not the physical conditions of the surrounding ambient medium. 

The line emission from these outflows is computed by post-processing the density and temperature profiles from \maihem\ using the microphysics code \cloudy\ \citep{Ferland2013}. 
Two sets of \cloudy\ models are run: one where \cloudy\ computes the ionization states assuming that the ionization states are in CIE, and one where the non-equilibrium ionization states are taken from \maihem.
We find that for the cooling, non-adiabatic conditions, the non-equilibrium atomic models tend to produce more highly ionized conditions compared to the equilibrium models.
This is true for nearly every element that is tracked by our atomics package. 
Therefore, we find that the non-equilibrium models tend to predict lower emissivities compared to the equilibrium models, for the nebular emission lines computed here.

Again using \cloudy, we have studied the effect of a photoionizing background on the line emission, using a SB99 ionizing SED appropriate for a SSC responsible for generating the modeled outflow.
Four models were considered: a control where no background UV field was applied, but using the density and temperature profiles from \maihem; one that considers pure photoionization on the same density profile, but assuming the gas is neutral, such that the temperature of the outflow is determined by photoheating only; one with density and temperature profiles from \maihem\ plus photoionization from the SB99 spectrum; and one with a \maihem\ model expanding into a lower density, 1 cm$^{-3}$ ambient medium at $10^4$~K.  The ambient density strongly affects the radial profile for the modeled emission.  
In high-density models, the collisional line emission is limited to the outer regions of the free-wind zone, and increases with radius; whereas in the low-density model, most collisional species decrease with radius except for a gap in the inner-most region.

The inclusion of a background UV field has little effect on high-ionization species generated by gas temperatures of $10^6 - 10^7$ K.
However, photoionization has a dramatic impact on some emission lines originating from a strongly cooling flow.  \CIV\ $\lambda$1549,1551  and \OVI\ $\lambda1037$, for example, show an increase in emission of roughly two orders of magnitude with photoionization, compared to without.  
In contrast, low-ionization species like [\OII\lbrack $\lambda3727$ are basically eliminated from the free-wind zone.  Moreover, in an optically thin, density-bounded shell, \HEII\ $\lambda1640$ and $\lambda4686$ are also strongly enhanced above photoionized values.  

Our models suggest that \CIV\ and \OVI\ may serve as diagnostics of catastrophic cooling conditions. Observations show that when seen as nebular line emission, these transitions are associated with extreme starbursts where catastrophic cooling is likely to occur. These include objects like extreme Green Peas, which are often found to be optically thin to LyC radiation.  \HEII, \CIII], \SiIV, and \SiIII\ may also be useful, especially where photoionization is density bounded.  Further study is needed to fully understand their emission.

Other evidence for suppressed superwinds and strong clumping in objects like extreme Green Peas is consistent with the presence of catastrophically cooling outflows, and the resulting picket-fence geometry can explain the escape of LyC radiation from these systems \citep{Jaskot2019}.
Although our 1-D models are not capable of simulating this clumping, the results presented here provide some initial insight on line emission from catastrophically cooling outflows, highlighting the importance of non-equilibrium atomic chemistry and predicted line diagnostics.

Understanding the expected line emissivities in these conditions can clarify the nature of Green Peas and their mechanism for LyC escape, as well as other starbursts experiencing these strongly cooling outflows.
Future studies will expand on the simulation conditions and study a wider parameter space that includes varying the mass input rate in order to gain a better understanding of when catastrophic cooling occurs and the range of predicted line emission.
Hydrodynamical effects, such as instability-induced gas clumping will also be studied by expanding these simulations into two dimensions.

\acknowledgments

We would like to thank the \cloudy\ community for useful discussions and suggestions.
Helpful comments by the referee are also gratefully acknowledged.  W.J.G. and M.S.O. acknowledge support from the University of Michigan, and M.S.O. also acknowledges NASA grant HST-GO-14080.002-A.  S.S. acknowledges support from CONACYT-M\'exico, research grant A1-S-28458; and E.S. was supported in part by the National Science Foundation under grant AST-1715876.
S.S and M.S.O. also thank the participants of the 2019 Guillermo Haro Workshop for friendly and helpful discussions.  M.S.O. thanks Robin Shelton  and Ash Danehkar for useful discussions, and Bob Benjamin for inspiration.
The software used in this work was in part developed by the DOE NNSA-ASC OASCR Flash Center at the University of Chicago.
The analysis presented here were created using the {\bf yt} analysis package \citep{Turk2011}. 

\software{\flash\ \citep{Fryxell2000}, yt \citep{Turk2011}, \cloudy\ \citep{Ferland2013}}
\bibliographystyle{apjsingle}
\bibliography{ms.bib}

\appendix

\section{Normalized Emission-Line Plots}
\label{apdx:LineEmission}
\begin{figure*}
\begin{center}
\includegraphics[trim=0.0mm 0.0mm 0.0mm 0.0mm, clip, width=0.85\textwidth]{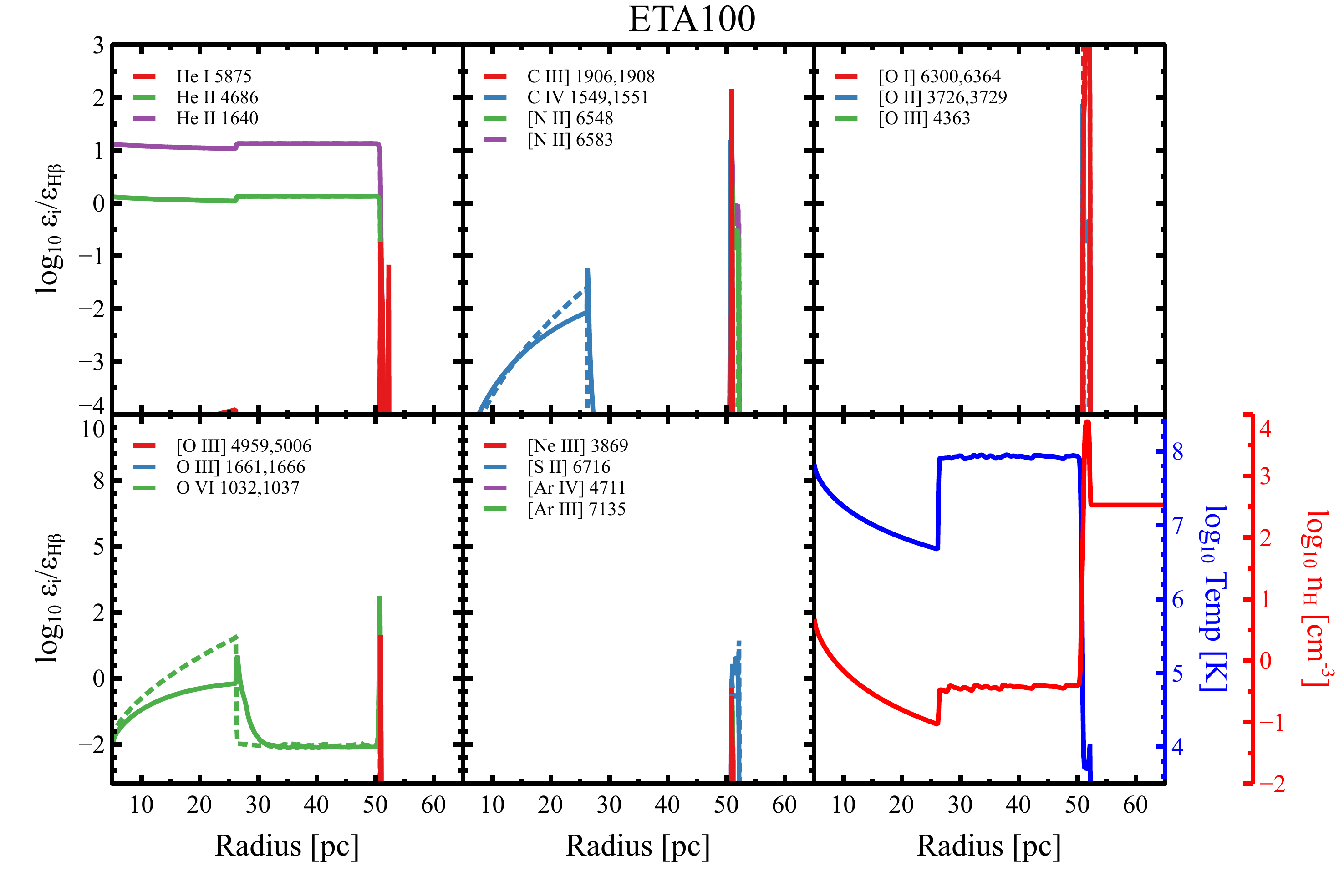}
\caption{ Line emission results for ETA100. This figure is the same as Figure~\ref{fig:R5V1000Lines} except now each line is normalized by H${\beta}$.  The solid lines represent emissivity ratios computed using the non-CIE, \maihem\ ionization states and the dotted line shows the equilibrium \cloudy\ results. }
\label{fig:R5V1000LinesAPDX}
\end{center}
\end{figure*}

\begin{figure*}
\begin{center}
\includegraphics[trim=0.0mm 0.0mm 0.0mm 0.0mm, clip, width=0.85\textwidth]{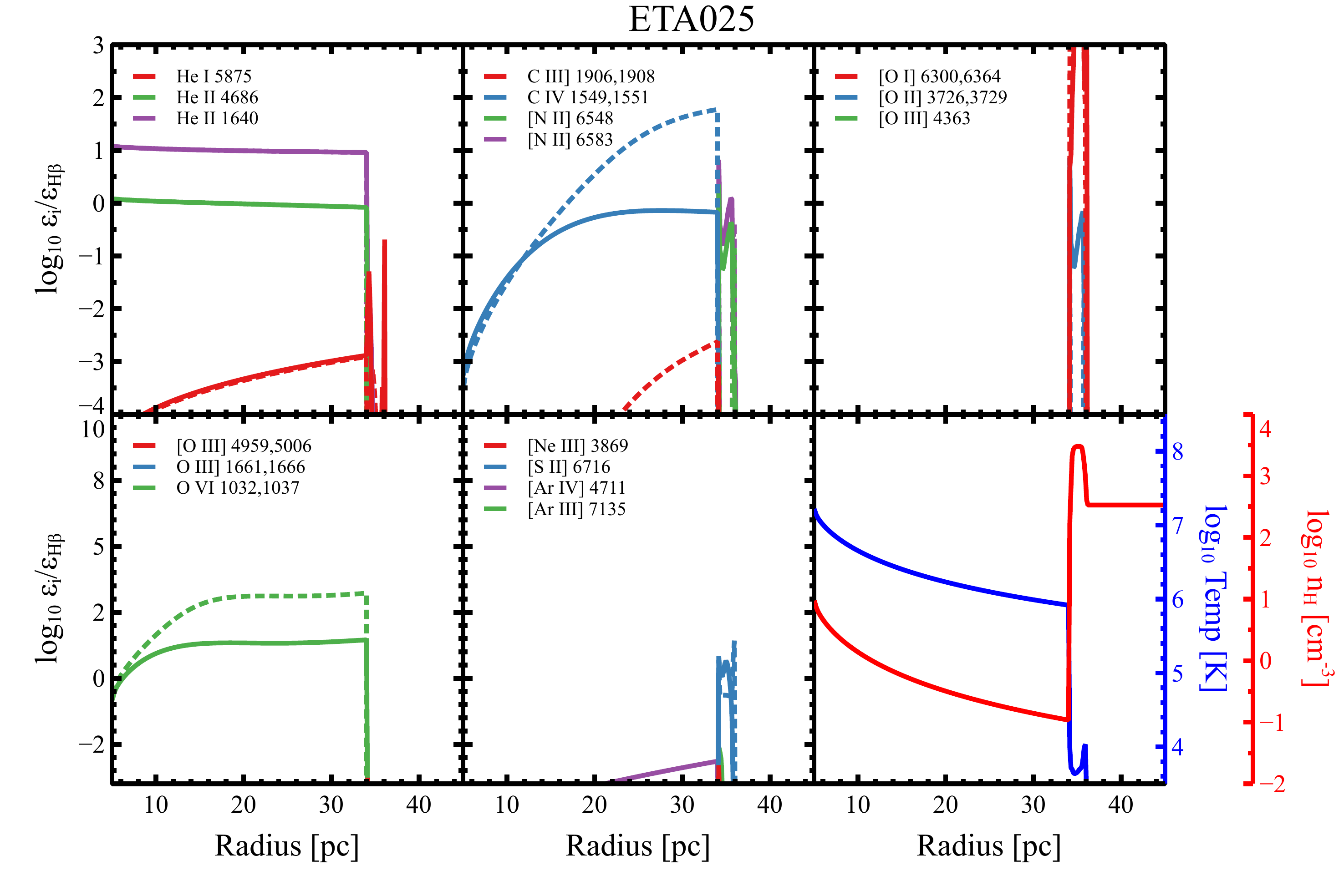}
\caption{ Line emission results for ETA025. This figure is the same as Figure~\ref{fig:R5V500Lines} except now each line is normalized by H${\beta}$.  }
\label{fig:R5V500LinesAPDX}
\end{center}
\end{figure*}

\begin{figure*}
\begin{center}
\includegraphics[trim=0.0mm 0.0mm 0.0mm 0.0mm, clip, width=0.85\textwidth]{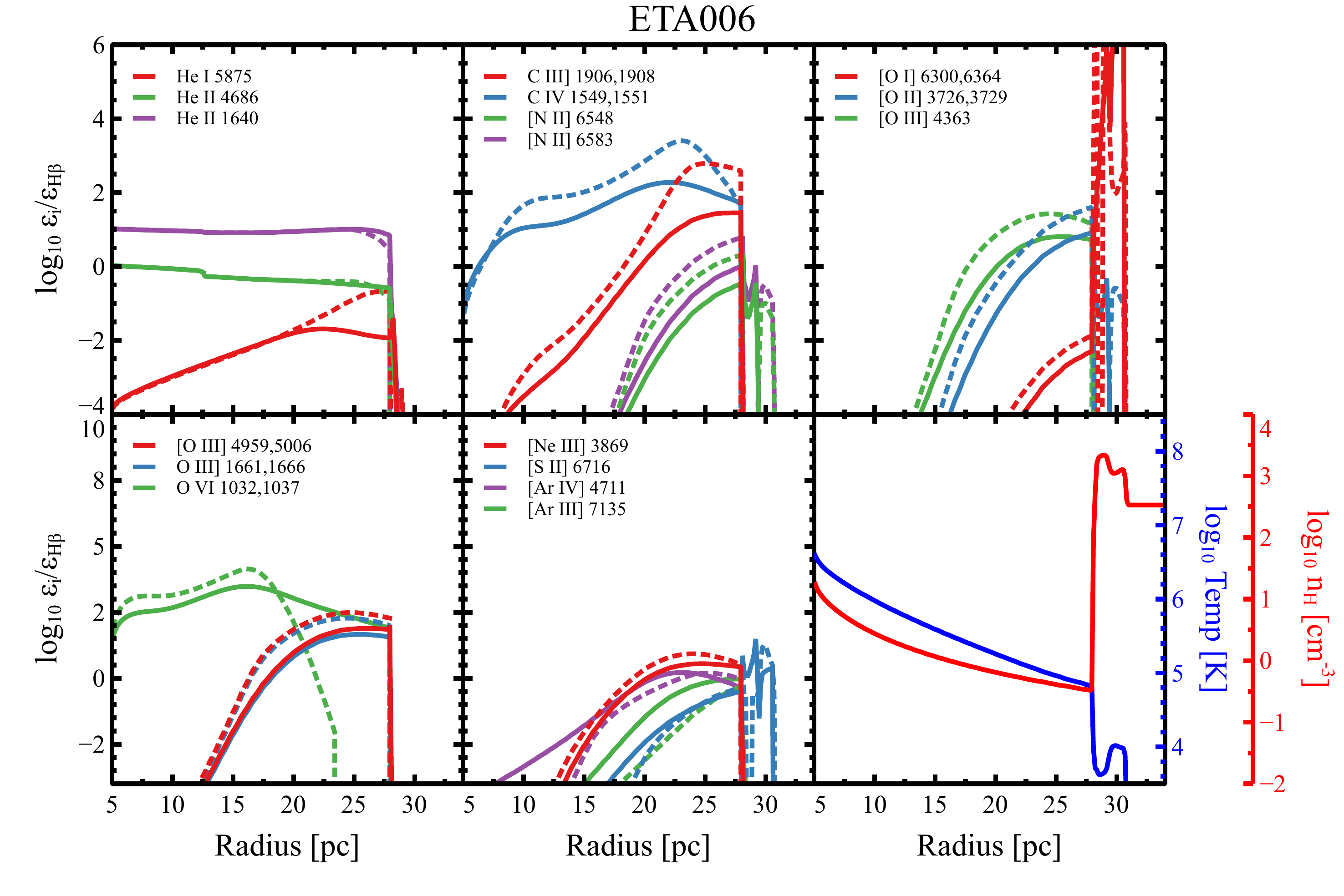}
\caption{ Line emission results for ETA006. This figure is the same as Figure~\ref{fig:R5V250Lines} except now each line is normalized by H${\beta}$. }
\label{fig:R5V250LinesAPDX}
\end{center}
\end{figure*}

\begin{figure*}
\begin{center}
\includegraphics[trim=0.0mm 0.0mm 0.0mm 0.0mm, clip, width=0.70\textwidth]{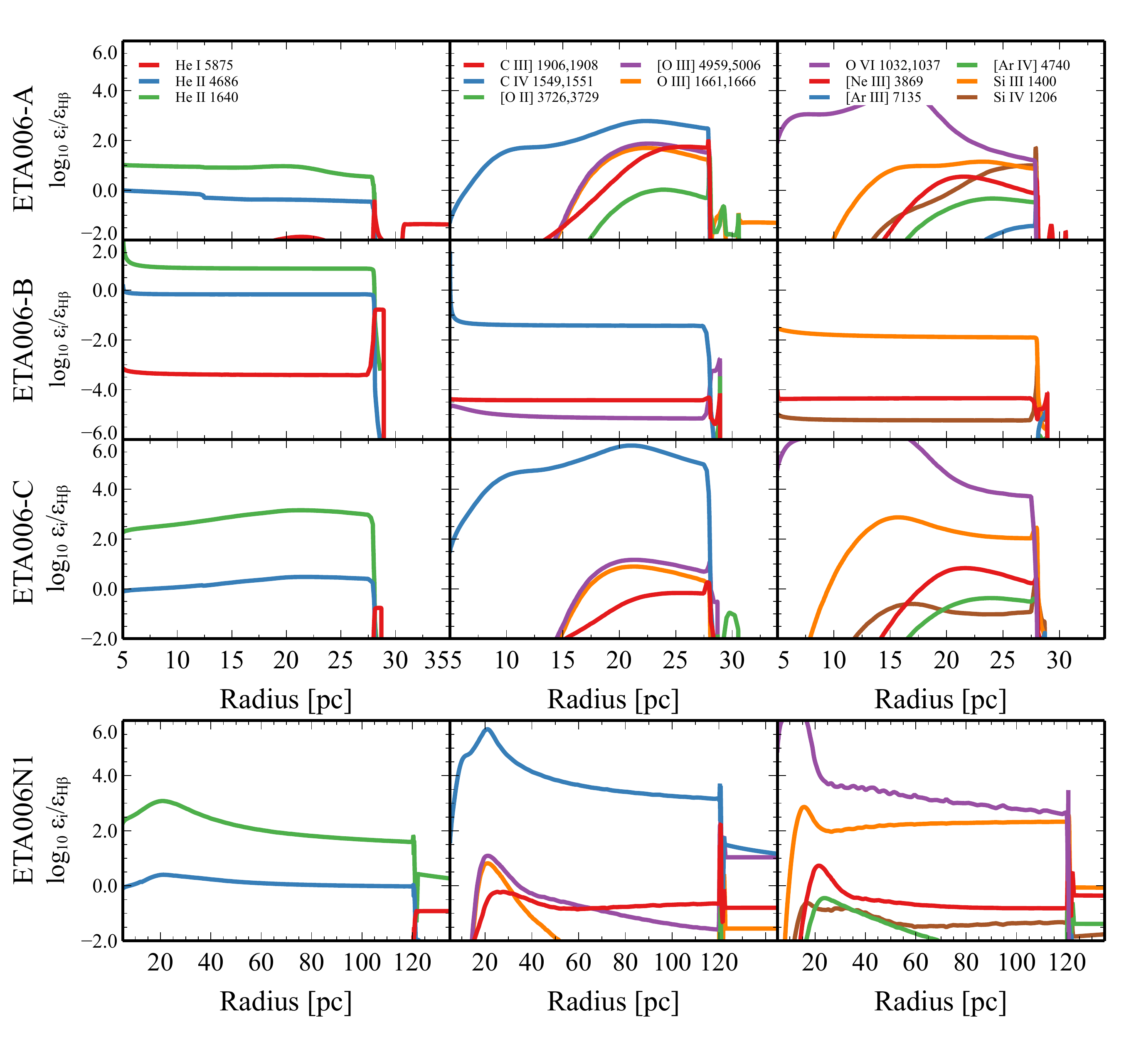}
\caption{ Results for ETA006 with photoionization. This figure is the same as Figure~\ref{fig:R5V250UV} except now each line is normalized by H${\beta}$.  }
\label{fig:R5V250UVLinesAPDX}
\end{center}
\end{figure*}

Here we plot the line emissivities relative to H${\beta}$ as a function of radius. 

\section{SSC Photoionization of Wind-Driven Shells}
\label{apdx:IonizingRadiation}
Wind-driven shells may be partially or completely ionized by the SSC Lyman continuum.
The gas density in the ionized part of the shell is:
\begin{equation}
\rho_{HII} = \mu_{i}P_{w}/kT_{HII}
\label{eqn:ap1}
\end{equation}
where $\mu_i$ = 14/23m$_{H}$ is the mean mass per particle in the completely ionized gas with 1 helium atom per each 10 hydrogen atoms, $m_H$ is the mass of the hydrogen atom, P$_{w}$ is the gas pressure in the shocked (model ETA100) or free (models ETA025 and ETA006) wind regions at the inner edge of the shell, $k$ is the Boltzmann constant and T$_{HII}$ = 10$^{4}$K is the ionized gas temperature. 
The ion number density then is:
\begin{equation}
n_i = \rho_{HII}/\mu_{a} = \mu_{i}P_{w}/k\mu_{a}T_{HII}
\label{eqn:ap2}
\end{equation}
where $\mu_a$= 14/11 m$_H$ is the mean mass per ion. 
The outer radius of the ionized zone R$_{HII}$ is determined by the equation:
\begin{equation}
Q = \frac{4\pi}{3}\beta n_i^2(R^3_{HII}-R^3_{in})
\label{eqn:ap3}
\end{equation}
where $Q$ is the number of the Lyman continuum photons emitted by the star cluster per unit time, R$_{in}$ is the shell inner radius and $\beta$= 2.59$\times$10$^{-13}$cm$^{3}$s$^{-1}$ is the hydrogen recombination coefficient to all but the ground level. 

The Starburst99 model and R$_{in}$ determine the value of $Q$ for our different models (see Fig.~\ref{fig:rhotemp}). 
Equations~\ref{eqn:ap1} and ~\ref{eqn:ap3} allow one to determine the ionized gas mass:
\begin{equation}
M_{HII} = \frac{4\pi}{3}(R_{HII}^3-R_{in}^3)\rho_{HII} = \frac{kQT_{HII}\mu^2_{a}}{\beta\mu_{i}P_{w}}
\label{eqn:ap4}
\end{equation}
One can compare this with the wind-driven shell mass:
\begin{equation}
M_{sh} = \frac{4\pi}{3}R^3_{sh}\rho_{amb}
\label{eqn:ap5}
\end{equation}
where R$_{sh}$ is the leading shock radius (see Fig.~\ref{fig:rhotemp}) and $\rho_{amb}$ is the ambient gas density (see Table~\ref{tab:runsummary}).
Note that in model ETA100 P$_w\sim$P$_{th}$ where P$_{th}$ is the gas thermal pressure in the shocked wind zone, whereas in models ETA025 and ETA006 P$_{w}\sim$ P$_{ram}$ where P$_{ram}$ = $\rho_{w}V^{2}_{w}$  is the star cluster wind
ram pressure at the inner edge of the shell. 
$\rho_{w}$ and V$_w\sim$2$\times v_{inj}$ are the star cluster wind density at the inner edge of the shell and the star cluster wind terminal speed, respectively.

\end{document}